\documentclass[a4paper,11pt]{article}

\usepackage{setspace}
\usepackage{amssymb}
\usepackage{amsmath}
\usepackage{mathtools}
\usepackage{booktabs}
\usepackage{enumerate}
\usepackage{graphicx}
\graphicspath{{Figures/}}
\usepackage{epsfig,amsmath,amssymb,graphicx,lscape,color}
\usepackage{pdflscape}
\usepackage{color}
\usepackage{float}
\usepackage{caption}
\usepackage{hyperref}
\usepackage{mathrsfs}
\usepackage{placeins}
\usepackage{xcolor}
\usepackage{subcaption}
\usepackage[export]{adjustbox}
\usepackage{mathdots}
\usepackage{calc}
\usepackage{array}

\usepackage{verbatim}
\usepackage{multirow}
\usepackage{tikz}
\usepackage{collcell}
\usepackage{xstring}
\usepackage{authblk}
\usepackage{bm}
\def\cbl{\color{black}}
\def\cb{\color{black}}
\DeclareMathOperator*{\argmax}{arg\,max}

 	\newenvironment{keywords}{\vspace{0.2cm}\footnotesize\begin{tabular}{rp{12cm}}\textbf{Keywords:} & }{\end{tabular}}
\begin{document}
\title{Parameter estimation and uncertainty quantification using information geometry}




\author[1,2,*]{Jesse A Sharp}
\author[1,2]{Alexander P Browning}
\author[1,2,3]{Kevin Burrage}
\author[1,4]{Matthew J Simpson}

\affil[1]{School of Mathematical Sciences, Queensland University of Technology, Brisbane, Australia}
\affil[2]{ARC Centre of Excellence for Mathematical and Statistical Frontiers, QUT, Australia}
\affil[3]{Visiting Professor, Department of Computer Science, University of Oxford, Oxford, UK}
\affil[4]{QUT Centre for Data Science, QUT, Australia}
\affil[*]{Corresponding author. E-mail: j3.sharp@qut.edu.au}


\date{\today}
\maketitle
\addtolength{\skip\footins}{2pc plus 5pt}
\renewcommand{\thefootnote}{\arabic{footnote}}


\renewcommand{\abstractname}{Abstract}
\begin{abstract}
	\noindent 
	In this work we: (1) review likelihood-based inference for parameter estimation and the construction of confidence regions; and, (2) explore the use of techniques from information geometry, including geodesic curves and Riemann scalar curvature, to supplement typical techniques for uncertainty quantification such as Bayesian methods, profile likelihood, asymptotic analysis and bootstrapping. These techniques from information geometry provide data-independent insights into uncertainty and identifiability, and can be used to inform data collection decisions. All code used in this work to implement the inference and information geometry techniques is available on \href{https://github.com/Jesse-Sharp/Sharp2021b}{GitHub}.
\end{abstract}





\section{Introduction} \label{sec:Introduction}

Computational and mathematical models are versatile tools, providing valuable insight into complex processes in the life sciences. Models can further our  understanding of mechanisms and processes, facilitate development and testing of hypotheses, guide experimentation and data collection and aid design of targeted interventions  \cite{Browning2021a,Murray2002,Sharp2020,Walpole2013,Warne2019a}. However, there are considerable challenges associated with calibrating these models to data. For example, models need to be sufficiently sophisticated to adequately reflect the behaviour of the underlying system, while ideally admitting \textit{identifiable} parameters that are interpretable, and that can be estimated from available or obtainable data \cite{Gabor2015,Liepe2014}. Further, available data can be limited and often is not collected for the express purpose of parameter estimation; data may be noisy, incomplete, or may not provide the level of detail or sample size required to obtain precise parameter estimates \cite{Frolich2014,Hlavacek2011,Schnoerr2017,Toni2009,Vo2015}. 

Due to the challenges associated with parameter estimation, we are often interested not only in point-estimates, but also the associated uncertainty \cite{DevenishNelson2010,Marino2008,Warne2017}. Quantifying and interpreting this uncertainty establishes a level of confidence in parameter estimates; and by extension, in the insights derived from the model. Further, this uncertainty quantification can give insights into \textit{identifiability}: whether the information in a data set can be used to infer unique or sufficiently precise parameter estimates for a given model \cite{Simpson2021}. Often we are concerned with both \textit{structural identifiability} and \textit{practical identifiability} \cite{Audoly2001,Hines2014,Simpson2020,Villaverde2016,Villaverde2019}. Structural identifiability can be thought of as a property of the underlying model structure and parameterisation; and refers to whether it is theoretically possible to determine unique parameter values, given an infinite amount of perfect noise-free data \cite{Browning2020,Simpson2021,Yeung2020}. Structural identifiability requires that unique parameter combinations precipitate distinct model outputs. Structural identifiability occurs if and only if the Fisher information matrix, which we soon discuss, is of full rank \cite{Komorowski2011}. In contrast, practical identifiability is less well defined, and depends on the quality and quantity of data available and existing knowledge of the parameters \cite{Browning2020}. In the context of profile likelihood methods, practical non-identifiability can manifest as contours of the log-likelihood function that do not admit closed levels; the log-likelihood does not reach a predetermined statistical threshold within the physical parameter regime \cite{Lill2019}. If a model is not structurally identifiable, it cannot be practically identifiable. 

\cbl Practical non-identifiability may be addressed either through improving data quantity or data quality \cite{Browning2020,Simpson2020}. Data quantity can be improved by increasing the number of observations; such as by making additional observations at different time points. Data quality may be improved through reducing the noise present in the data, for example by obtaining a dataset with reduced measurement error or repeating measurements across identically-prepared experiments \cite{Brouwer2016,Raue2009}. It is also possible to resolve practical non-identifiability through incorporating existing knowledge about parameters, such as physical constraints or information established in previous studies; or specifically in the Bayesian inferential framework, through informative priors \cite{Gelman1996}. Addressing structural non-identifiability is more challenging, for example this may necessitate a change to the underlying model structure \cite{Godfrey1980,Raue2009,Villaverde2016}.\cb

Uncertainty quantification takes many forms, with common examples including Bayesian methods, profile likelihood, asymptotic analysis and bootstrapping \cite{Frolich2014,Gelman1995,Mitra2019,Vo2015,Warne2019}. Bayesian methods are widely used for parameter estimation and uncertainty quantification, with Bayesian computation being employed throughout the mathematical biology and systems biology literature. Broadly, these methods involve repeated sampling of parameter values from a prior distribution and invoking Bayes theorem to approximate the \textit{posterior} distribution; the posterior distribution describes knowledge about the probability of parameter combinations after taking into account the observed data and any prior information \cite{Browning2020,Warne2019}. \cbl Well-known approaches include rejection sampling, Markov Chain Monte Carlo (MCMC) and sequential Monte Carlo (SMC) or particle filtering. In rejection sampling, parameters drawn from a prior distribution are used to simulate the model; simulated data is compared to the observed data based on some distance metric, and if this metric is within a prescribed tolerance, the parameters are accepted as a sample from the approximate posterior distribution, otherwise they are rejected \cite{Gelman1995,Lambert2018b}. Rejection sampling can be computationally expensive as the rejection rate can be significant with an uninformative prior \cite{Liepe2010,Prangle2017}. In MCMC the parameter space is sampled following a Markov chain---a memoryless stochastic process where the probability of the next state depends only on the previous state \cite{Pawitan2001}---with a stationary distribution corresponding to the posterior distribution. Samples are accepted or rejected based on the relative likelihood between the current configuration and proposed sample \cite{Andrieu2003,Liu1998,Toni2009,Warne2019}. For SMC, rejection sampling can be used to produce an initial coarse approximation of the posterior distribution. This coarse approximation informs further (sequential) sampling efforts in the region of parameter space corresponding to high likelihood, reducing the rejection rate when compared to rejection sampling alone \cite{Jabot2013,Liepe2010,Toni2009}. MCMC and SMC approaches can offer significantly improved efficiency in comparison to rejection sampling \cite{Liepe2010,Warne2019}, but both involve specifying hyperparameters and these choices are not always obvious.\cb 

 \cbl In situations where the likelihood function is intractable or not easily evaluated, Approximate Bayesian Computation (ABC) provides a range of related likelihood-free methods for estimating posterior distributions \cite{Sunnaker2013}. Popular approaches include ABC rejection sampling \cite{Jabot2013,Kursawe2018,Prangle2017,Sunnaker2013,Wilkinson2013}, ABC MCMC \cite{Marjoram2003,Siekmann2012,Sisson2018}, and ABC SMC \cite{Liepe2010,Toni2009}; we do not focus on ABC methods here, \cbl as the approaches we explore in this work are applied to problems with tractable likelihoods. \cb We direct interested readers to the wealth of information in the references provided. 

For Bayesian inference methods, uncertainty can be quantified based on features such as the coefficient of variation and probability intervals of the posterior distribution \cite{Vo2015}. There are a variety of approaches for uncertainty quantification for frequentist inference methods. In profile likelihood, a parameter of interest is varied over a fixed set of values, while re-estimating the other parameters; providing insight into identifiability and uncertainty \cite{Browning2021a}. In asymptotic analysis, confidence regions can be constructed based on local information via a Taylor expansion of the Fisher information about the maximum likelihood estimate (MLE) \cite{Frolich2014,Lill2019}. In bootstrapping, data is repeatedly sampled and parameter estimates are computed from the samples; these estimates are used to construct confidence intervals \cite{Mitra2019}.   

Through the geometric approaches we review in this work, more akin to traditional approaches for sensitivity analysis \cite{Marino2008,Miao2011,Yue2006}, we explore the curvature of the parameter space through an information metric induced by the likelihood function. Whereas likelihood-based approximate confidence regions provide insight into specific level curves of the likelihood function---the levels of which depend on an asymptotic large sample argument \cite{Pawitan2001}---this geometric approach provides insight into the \textit{shape} and sensitivity of the parameter space. For example, we compute \textit{geodesic curves} that describe the geometric relationship between distributions with different parameters \cite{Menendez1995}; and explore the \textit{scalar curvature} throughout parameter spaces. We review ideas from \textit{information geometry} in the context of inference and uncertainty quantification; not with a view to replacing established methods such as profile likelihood, asymptotic analysis, bootstrapping and Bayesian methods \cite{Frolich2014,Mitra2019,Vo2015,Warne2019}, but rather to supplement them where additional insight may prove useful. 

Information geometry is a branch of mathematics connecting aspects of information theory including probability theory and statistics with concepts and techniques in differential geometry \cite{Amari1985}. In this exposition we seek to outline only the key concepts required to understand the information geometric analysis in this work. However, we note that more thorough and rigorous treatments of the concepts introduced in this section, and mathematical foundations of information geometry, can be found in texts and surveys such as \cite{Amari1985,Calin2014,Nielsen2020}. Central to the information geometry ideas explored in this work is the concept of a \textit{statistical manifold}; an abstract geometric representation of a distribution space, or a Riemannian manifold consisting of points that correspond to probability distributions, with properties that we later discuss. For example, the set of normal distributions parameterised by mean, $\mu$, and standard deviation, $\sigma>0$:
\begin{align}
p(x;\mu,\sigma) = \frac{1}{\sigma\sqrt{2\pi}}\exp\left[{-\frac{(x-\mu)^2}{2\sigma^2}}\right], \quad x\in\mathbb{R}, \label{eq:UnivNormal}
\end{align}     
can be thought of as a two-dimensional surface with coordinates  $(\mu,\sigma)$ \cite{Calin2014}. In this work we will use $\boldsymbol{\theta}$ to refer to the parameters of interest that we seek to estimate; i.e. $\boldsymbol{\theta} = (\mu,\sigma)$ for the univariate normal distribution with unknown mean and standard deviation. In Section \ref{sec:Results}, we consider various combinations of components of $\boldsymbol{\theta}$; including model parameters, variability in observations characterised by a separate observation noise model, and initial conditions associated with a differential equation-based process model. When referring to all possible parameters, rather than solely the unknown parameters to be estimated, we denote this $\boldsymbol{\Theta}$.  

In applications where we consider multiple data sets, or different candidate models or candidate parameter values, we are interested in methods of comparing distributions. A well-known measure for comparing a probability distribution, $P$, to another, $Q$, is the Kullback-Leibler (KL) divergence from $P$ to $Q$, denoted $\mathcal{D}_{KL}(P,Q)$ \cite{Eguchi2006}. The KL divergence, or \textit{relative entropy}, can be computed as \cite{Eguchi2006}: 
\begin{align}\mathcal{D}_{KL}(P,Q) = \int p(x) \log \frac{p(x)}{q(x)} \textrm{d}x = \mathbb{E}_p \left[\log \frac{p(x)}{q(x)}\right],\label{eq:KLdivergence}
\end{align} 
where $p(x)$ and $q(x)$ are the respective probability density functions of $P$ and $Q$. Consider two sets of parameters, $\boldsymbol{\theta}^*$ and $\hat{\boldsymbol{\theta}}$; let $\log(p(x)) = \log(p(x|\boldsymbol{\theta}^*)) = \ell(\boldsymbol{\theta}^*)$ and $\log(q(x)) = \log(p(x | \hat{\boldsymbol{\theta}})) = \ell(\hat{\boldsymbol{\theta}})$, where $\ell(\cdot)$ denotes the log-likelihood, discussed in detail in Section \ref{sec:Methods}. If $p(x|\boldsymbol{\theta}^*)$ is the true distribution and $p(x | \hat{\boldsymbol{\theta}})$ is our estimate, then (\ref{eq:KLdivergence}) is the expected log-likelihood ratio and the relationship between MLE and KL divergence becomes evident; maximising the likelihood is equivalent to minimising KL divergence \cite{Murphy2012}.  

An issue with the KL divergence is asymmetry; $\mathcal{D}_{KL}(P,Q) \ne \mathcal{D}_{KL}(Q,P)$. It is not necessarily obvious in a given situation which orientation of the KL divergence will most appropriately inform decisions such as model selection \cite{Seghouane2007}. Due to the aforementioned asymmetry, and its failure to satisfy the triangle inequality, the KL divergence is not a \textit{metric}---it is not a measure of distance in a differential geometric sense---on a given manifold \cite{Calin2014}. One means of addressing this asymmetry is through devising various symmetrised forms of the KL divergence to inform model selection criteria \cite{Seghouane2007}. Alternatively, we may approach the issue from a geometric perspective.          
It is natural to think of geometry in terms of objects or shapes in Euclidean, or \textit{flat}, space. Euclidean space is characterised by orthonormal basis vectors; the standard basis in three-dimensions being $\bm{e}_1 = (1,0,0)^\textrm{T}$, $\bm{e}_2 = (0,1,0)^\textrm{T}$, $\bm{e}_3 = (0,0,1)^\textrm{T}$, where superscript $\textrm{T}$ denotes the transpose. In the $n$-dimensional orthonormal basis, we can compute the squared infinitesimal distance between the points $\mathcal{S}$ and $\mathcal{S} + \textrm{d}s$, where $\textrm{d}s$ has components $\textrm{d}s_i$, as  \cite{Amari1998}:
\begin{align} 
||\textrm{d}s||^2 = \sum_{i=1}^{n}(\textrm{d}s_i)^2.\label{eq:OrthLocalDistance}
\end{align}
Differential geometry extends ideas from Euclidean geometry to manifolds. Manifolds are topological spaces that resemble flat space about each individual point in the space; they can be considered \textit{locally flat}, but have a different topology globally. The sphere is a classic example, whereby points on the surface are locally topologically equivalent to two-dimensional Euclidean space, but globally the sphere is curved and has a compact topology; it is bounded and closed \cite{Amari2016}. In particular, we are interested in Riemannian manifolds; differentiable manifolds---sufficiently locally smooth that our typical notions of calculus remain valid---upon which we are able to measure geometric quantities such as distance, through the existence of a Riemannian metric on the tangent space of the manifold, that generalises the notion of an inner product from Euclidean geometry \cite{Lee2018}.

A Riemannian metric is a smooth covariant 2-tensor field; on a differentiable manifold $M$, the Riemannian metric is given by an inner product on the tangent space of the manifold, $T_pM$, which depends smoothly on the base point $p$ \cite{Jost2017,Lee2018}. \cbl A tangent space can be thought of as a multidimensional generalisation of a tangent plane to a three-dimensional surface. \cb  Each point $p$ on a manifold is associated with a distinct tangent space. An $n$-dimensional manifold has infinitely many $n$-dimensional tangent spaces; the collection of these tangent spaces is referred to as the tangent bundle of the manifold. On a manifold each tangent space can have different basis vectors, in contrast to Euclidean geometry where tangent vectors at any point have the same basis vectors. A consequence of the distinct basis vectors of tangent spaces on manifolds is that tangent vectors at different points on the manifold cannot be directly added or subtracted. Introducing an \textit{affine connection} on the manifold connects nearby tangent spaces, such that the manifold looks infinitesimally like Euclidean space, facilitates differentiation of tangent vectors \cite{Amari2000}. Formally, we introduce the unique, torsion free Levi-Civita connection, $\nabla$; an affine connection on the Riemannian manifold that yields isometric parallel transport, such that inner products between tangent vectors, defined by the metric, are preserved \cite{Giesel2021}. The coefficients of this connection are the Christoffel symbols, which we discuss further in Section \ref{sec:Methods}. Readers are directed to \cite{Amari2000,Arwini2008,Giesel2021} for further detail regarding the Levi-Civita connection, and how it relates to other concepts discussed in this work. A manifold equipped with such a connection and a Riemann metric is a Riemann manifold.

Metric tensors can be thought of as functions that facilitate computation of quantities of interest such as distances on a manifold. A metric matrix with elements $g_{ij}$, $G = [g_{ij}]$, is positive definite and symmetric \cite{Lee2018}. The metric matrix defines an inner product between $u$ and $v$ as $\langle {u,v} \rangle_G = u^\textrm{T}Gv$, and the length of $u$ as $||u||_G = \sqrt{\langle u,u \rangle_G}$ \cite{Costa2015}. On a Riemannian manifold we consider a generalisation of the square of the infinitesimal distance element (\ref{eq:OrthLocalDistance}), appropriate for non-orthonormal bases \cite{Amari1998}, given by 
\begin{align}
||\textrm{d}s||^2 = \sum_{i,j = 1}^{n} g_{ij}\textrm{d}s_i\textrm{d}s_j.\nonumber
\end{align}

Foundational to information geometry is the notion that the Fisher information matrix defines a Riemannian metric on a statistical manifold \cite{Rao1992}. The Fisher information, denoted $\mathcal{I}(\bm\theta)$, describes the expected curvature of the log-likelihood and gives information about the precision and variance of parameter estimates. Therefore, $\mathcal{I}(\bm\theta)$  can incorporate information about both the curvature induced by the data through the observation process, as well as the curvature induced by parameter sensitivities through a mathematical model that links parameter estimates to data. In the examples we consider, deterministic model predictions are connected to the data through the probabilistic observation process, yielding a general formula for the Fisher information \cite{Lehmann1998}:
\begin{equation}\label{eq:FIM}
\mathcal{I}(\bm\theta) = N\mathbf{J}(\bm\theta)^T %
\overbrace{\vphantom{x^T}\mathcal{O}(\mathbf{m})}^%
{\mathclap{\substack{\text{Curvature induced}\\\text{by data}}}}%
\underbrace{\vphantom{x^T}\mathbf{J}(\bm\theta)}_%
{\mathclap{\substack{\text{Curvature induced}\\\text{by parameter sensitivities}}}}.
\end{equation}	
Here, we denote $\mathcal{O}(\mathbf{m})$ the Fisher information matrix of the observation process, given a model, $\mathbf{m} = \mathbf{m}(\bm\theta)$, where $\mathbf{J}(\bm\theta)$ is the Jacobian of the model with respect to the parameters. The number of independent, identically distributed (iid) observations in the likelihood is given by $N$; with statistical independence, the Fisher information is additive \cite{Frieden1990}. 

Expression (\ref{eq:FIM}) highlights a link between sensitivity analysis, structural identifiability and practical identifiability \cite{Maclaren2020}. For sensitivity analysis and structural identifiability, only the curvature of the model space is studied through $\mathbf{J}(\bm\theta)$. In practical identifiability analysis, the sensitivity of the model is linked to the data through an observation process, and the curvature of the parameter space is studied through, for example, $\mathcal{I}(\bm\theta)$. 

In this review, we present and explore fundamental techniques in inference and information geometry, including confidence regions, geodesic curves, and scalar curvature. Through application to standard distributions and canonical models in the life sciences, including population growth processes and epidemic transmission, we demonstrate how these techniques can be combined to provide additional insights into parameter estimation and uncertainty quantification. Starting with parameter estimates inferred from real data, we use mathematical models to generate synthetic data with different numbers of observations and at varying points in time, to explore the impact that these aspects have on the inference and information geometry results. Specifically, we consider univariate and multivariate normally distributed observation processes; linear, exponential and logistic models of population growth; and the classical susceptible, infectious, recovered (SIR) model of epidemic transmission \cite{EdelsteinKeshet2005,Kermack1927}. Although the examples considered in this work are based on ODE process models drawn from the life sciences, the techniques we consider are general and can be applied in the context of parameter estimation and uncertainty quantification in any discipline and for other model formulations.  

By considering standard distributions and canonical models we are able to explore the inference and information geometry techniques through a series of examples with incremental increases in complexity. Through this approach, we consider the techniques as applied to both linear and nonlinear ordinary differential equation (ODE) models, coupled nonlinear ODE systems, and data with both one and many observed variables. We consider cases where model parameters, initial conditions, and the standard deviation of the data, are to be estimated from data. The inference and information geometry techniques considered in this work are general, and can be applied far more widely than the examples we consider here. To improve the accessibility of these methods, code used to implement the inference and information geometry techniques applied in this work is written in the open source Julia language \cite{Bezanson2012}; and is available on \href{https://github.com/Jesse-Sharp/Sharp2021b}{GitHub}.

In Section \ref{sec:Methods} we describe the inference and information geometry methods implemented in this work, including maximum likelihood estimation, profile-likelihood based approaches, geodesic curves and scalar curvature calculations. Results of applying these techniques to univariate and multivariate normal distributions, linear, exponential and logistic growth models and the SIR model, are presented in Section \ref{sec:Results}. We discuss the utility of these techniques, and identify opportunities for extension and further consideration in Section \ref{sec:Discussion}.

\section{Methods}\label{sec:Methods}

Here we describe the parameter inference and information geometry methods used to produce results in this work. We also describe the numerical methods used to implement these techniques. The techniques we discuss in this section readily generalise to parameter spaces with an arbitrary number of dimensions, so we discuss the techniques here for arbitrary dimensions. However, for the sake of exploring the techniques through visualisation in Section \ref{sec:Results}, we restrict ourselves to two-dimensional manifolds. In context, this means we consider only two parameters to be inferred in any given example, treating other parameters as known and fixed; for example, as if they are drawn from prior knowledge or pre-estimated.

Although we consider deterministic mathematical models, data used to estimate parameters can exhibit significant variability. We follow a standard approach and assume that the mathematical models describe the expected behaviour, and that our observations are normally distributed about this expected behaviour \cite{Hines2014}. This allows us to think about a statistical model, $\boldsymbol{m}(\boldsymbol{\theta},t)$, in terms of its expected behaviour, $\boldsymbol{\mu}$ and the standard deviation of the observations, $\sigma$. 

\begin{align*}
\boldsymbol{m}(\boldsymbol{\theta},t) = (\boldsymbol{\mu}(\boldsymbol{\theta},t),\sigma(\boldsymbol{\theta},t)).
\end{align*} 

We restrict the examples in this work to cases where $\sigma$ is constant; setting $\sigma(\boldsymbol{\theta},t) =\sigma$. In this work we focus on the most commonly employed additive noise model \cite{Hines2014,Raue2009,Simpson2020,Toni2009,Warne2019a}. Additive noise implies that the variance of the data is independent of the mean behaviour. In cases where variance scales with mean behaviour, multiplicative noise may be more appropriate. The information geometric methods presented here are applicable in cases where the Fisher information can be obtained; including models with multiplicative noise, and parameter or time dependent standard deviation. However, obtaining the Fisher information is a separate challenge, and can be difficult when considering different process and noise models. 

\subsection{Parameter inference}

In this work, parameter estimates are inferred from data following a standard maximum log-likelihood based approach. We make observations at $L$ time-points, $T = (t_1,t_2,...,t_L)$. At each time-point we make $N$ observations, $\mathcal{X} =(\mathbf{x}_1(T),\mathbf{x}_2(T),...,\mathbf{x}_N(T))$. With this notation the log-likelihood function is
\begin{equation}
\ell(\bm\theta;\mathcal{X}) = \sum_{j=1}^L\sum_{i=1}^{N}  \log f\left(\mathbf{x}_i(t_j);\boldsymbol{\mu}(\bm\theta,t_j),\sigma^2\right),\label{eq:llfun} 
\end{equation}
where $f(\mathbf{x};\bm\mu,\sigma^2)$ is the probability density function associated with our observation process. In this work we hold $N$ constant across time-points, though non-constant $N$ is easily incorporated into Equation (\ref{eq:llfun}) as $N_j$.   The likelihood function can be thought of as the joint probability density of all the data for a given set of parameters. In examples where $\sigma$ is unknown, we treat $\sigma$ as an element of $\boldsymbol{\theta}$, but note that the expected model behaviour is independent of $\sigma$.  The MLE is the point estimate, $\hat{\boldsymbol{\theta}}$, that satisfies
\begin{equation}\label{eq:MaximumLikelihood}
\hat{\bm\theta} = \underset{\bm\theta}{\argmax} \:\ell(\bm\theta;\mathcal{X}),
\end{equation}
where $\argmax(\cdot)$ returns the argument, $\boldsymbol{\theta}$, that maximises $\ell(\bm\theta;\mathcal{X})$ in (\ref{eq:MaximumLikelihood}). The associated maximum log-likelihood is $\ell(\boldsymbol{\hat{\theta}})$.  MLEs of the parameters of interest are obtained by solving (\ref{eq:MaximumLikelihood}) numerically as outlined later in Section \ref{sec:Methods}.  For an iid sample from a univariate normal distribution, $\mathcal{N}(\mu,\sigma^2)$, maximising the likelihood function of $\mu$ is equivalent to performing least-squares estimation \cite{Browning2020}, although having access to the likelihood function facilitates uncertainty quantification. 

Presenting confidence regions alongside MLEs enhances our interpretation of the likelihood function, while still acknowledging that the estimates carry uncertainty \cite{Pawitan2001}. We apply a probability-based log-likelihood approach when constructing confidence regions for model parameters. \cbl From Wilks' theorem \cite{Pawitan2001}, asymptotically as $N \rightarrow \infty$, \cb  an approximate $\alpha$-level confidence region is given by
\begin{equation}
\left\{ \boldsymbol{\theta} : \ell(\boldsymbol{\theta}) \ge \ell(\hat{\boldsymbol\theta}) - \frac{\Delta_{\nu,\alpha}}{2}\right\}, \label{eq:LikelihoodConfRegion}
\end{equation}
where $\Delta_{\nu,\alpha}$ is the $\alpha$th-quantile of the $\chi^2(\nu)$ distribution; with $\nu$ degrees of freedom \cite{Browning2021a}. In this work the degrees of freedom correspond to the number of parameters of interest, i.e. $\nu = \textrm{dim}(\boldsymbol{\theta})$. To enable comparison between different data sets and models in Section \ref{sec:Results}, we consider the normalised log-likelihood, $\hat{\ell}(\boldsymbol{\theta})= \ell(\boldsymbol{\theta}) - \ell(\boldsymbol{\hat{\theta}})$. This forms the basis for log-likelihood ratio based hypothesis tests \cite{Pawitan2001}. The normalised log-likelihood is zero at the MLE: $\hat{\ell}(\boldsymbol{\hat{\theta}}) \equiv 0$. 

\subsection{Information geometry}

As outlined in Section \ref{sec:Introduction}, the Fisher information describes the curvature of the log-likelihood function. It describes how much information a random variable, $X$, contains about a parameter, $\boldsymbol{\theta}$. For unbiased estimators, the inverse of the Fisher information provides a lower bound on the covariance matrix, via the Cramer-Rao inequality \cite{Watanabe2009}. Formally, the Fisher information is the covariance of the score, where the score is defined as the partial derivative of the log-likelihood with respect to $\boldsymbol{\theta}$ \cite{Lehmann1998,Pawitan2001}. The Fisher information matrix can be written as \cite{Kay1993,Pawitan2001}:
\begin{align}
[\mathcal{I}(\boldsymbol{\theta})]_{ij} = \mathbb{E}_{X}\left[\left(\frac{\partial}{\partial \theta_i}\log f(X;\boldsymbol{\theta})\right)\left(\frac{\partial}{\partial \theta_j}\log f(X;\boldsymbol{\theta})\right) \right]. \label{eq:FisherDef}
\end{align} 
\cbl We can recover our expression for the Fisher information in Equation (\ref{eq:FIM}) from Equation (\ref{eq:FisherDef}), by considering how Equation (\ref{eq:FisherDef}) changes under reparameterisation, and via application of the chain-rule for differentiation \cite{Lehmann1998}. \cb
  
With observations at $L$ unique times, $T = (t_1, t_2, ...,t_L)$, we can think of a model as a mapping between the parameters and the outputs that we can observe: 
\begin{align} 
\mathbf{m}(\boldsymbol{\theta}) : \boldsymbol{\theta} \to \bigg(\big(\boldsymbol{\mu}_1(\boldsymbol{\theta},t_1),\sigma\big),\big(\boldsymbol{\mu}_2(\boldsymbol{\theta},t_2),\sigma\big),...,\big(\boldsymbol{\mu}_L(\boldsymbol{\theta},t_L),\sigma\big)\bigg). \label{eq:modelmapping}
\end{align} 

We consider some examples where $\sigma$ is unknown and is estimated as a part of the analysis; in these instances $\sigma \in \boldsymbol{\theta}$, however we express $\sigma$ explicitly in the mapping presented in  (\ref{eq:modelmapping}) to emphasise that it behaves somewhat differently to a model parameter. The expected behaviour of the model does not depend on $\sigma$, and variability in the data maps directly to $\sigma$. In all the examples we consider, $\sigma$ is constant. This could be extended to incorporate variability dependent on the expected behaviour, for example logistic growth with standard deviation that depends on the population density \cite{Browning2021b}. In the mapping, this could be expressed as $\sigma(\boldsymbol{\mu}(\boldsymbol{\theta},t))$.     

Following Equation (\ref{eq:FIM}), we can form the Fisher information as a combination of the Fisher information matrix of the observation process, $\mathcal{O}(\mathbf{m})$, and the Jacobian of the model with respect to the parameters, $\mathbf{J}(\boldsymbol{\theta})$. From (\ref{eq:modelmapping}), with $\nu$ unknown parameters ($\dim(\boldsymbol{\theta})=\nu$), we can view the model Jacobian as
\begin{equation}
\mathbf{J}(\boldsymbol{\theta}) = \begin{pmatrix} 
\frac{\partial \boldsymbol{\mu}_1}{\partial \theta_1} &\frac{\partial \boldsymbol{\mu}_1}{\partial \theta_2}&\dots &\frac{\partial \boldsymbol{\mu}_1}{\partial \theta_\nu}\\ \frac{\partial \sigma}{\partial \theta_1} & \frac{\partial \sigma}{\partial \theta_2}&\dots &\frac{\partial \mathbf{\sigma}}{\partial \theta_\nu}\\\frac{\partial \boldsymbol{\mu}_2}{\partial \theta_1} &\frac{\partial \boldsymbol{\mu}_2}{\partial \theta_2}&\dots &\frac{\partial \boldsymbol{\mu}_2}{\partial \theta_\nu}\\ \frac{\partial \sigma}{\partial \theta_1} & \frac{\partial \sigma}{\partial \theta_2}&\dots&\frac{\partial \mathbf{\sigma}}{\partial \theta_\nu}\\ \vdots& \vdots&&\vdots\\
\frac{\partial \boldsymbol{\mu}_j}{\partial \theta_1} &\frac{\partial \boldsymbol{\mu}_j}{\partial \theta_2}&\dots &\frac{\partial \boldsymbol{\mu}_j}{\partial \theta_\nu}\\
\frac{\partial \sigma}{\partial \theta_1} & \frac{\partial \sigma}{\partial \theta_2}&\dots&\frac{\partial \mathbf{\sigma}}{\partial \theta_\nu}
\end{pmatrix}.\label{eq:modelJAC}
\end{equation}

Noting that we are taking $\sigma$ to be independent of model parameters, all of the partial derivatives of $\sigma$ in (\ref{eq:modelJAC}) are zero, except the case where $\theta_i = \sigma$, for some $i \in \{1,2,...,\nu\}$,  whereby the corresponding partial derivative is unity. Given a set of $N$ normally distributed observations at a single point in time, we have an observation process characterised by a mean, $\mu$, and standard deviation, $\sigma$. The Fisher information for such an observation is given by 
\begin{equation}
\mathcal{I}(\mu,\sigma) = \frac{N}{\sigma^2} \mathbf{D}, \quad\textrm{where } \mathbf{D} = \textrm{diag}(1,2).  \label{eq:normalFIM}
\end{equation}

This can be verified by applying Equation (\ref{eq:FisherDef}) to (\ref{eq:UnivNormal}). For data at $L$ time-points with $N_1, N_2, ..., N_L$ observations at each time, with constant standard deviation, the Fisher information for the observation process is a $2L\times2L$ (block) diagonal matrix: 
\begin{equation} 
\mathcal{I}(\mu,\sigma) = \textrm{diag}\left(\frac{N_1}{\sigma^2} \mathbf{D},\frac{N_2}{\sigma^2} \mathbf{D},...,\frac{N_L}{\sigma^2} \mathbf{D}\right).\label{eq:FIM2Nx2N}
\end{equation}

Similarly, for a model with $M$ species, where we have observations of all $M$ species at only one time-point we recover Fisher information in the form of (\ref{eq:FIM2Nx2N}). For observations of $M$ species at $L$ time-points we form a $2LM\times2LM$ (block) diagonal matrix from (\ref{eq:FIM2Nx2N}). Assuming a constant standard deviation, for the computations in this work we could more simply express (\ref{eq:FIM2Nx2N}) as the diagonal matrix $\textrm{diag}(N_1/\sigma^2,N_2/\sigma^2,...,N_L/\sigma^2,2\sum N_i/\sigma^2)$, where $\sum N_i$ is the total number of observations contributing to our information regarding the standard deviation, and the factor of two comes from (\ref{eq:normalFIM}). In this case, the model Jacobian as presented in (\ref{eq:modelJAC}) is modified such that only the final row includes the partial derivatives with respect to the standard deviation. 

\cbl Before outlining specific techniques of information geometry, we present a conceptual example to develop some intuition for information geometric concepts. Consider the manifold corresponding to the family of univariate normal distributions parameterised by mean, $\mu$, and standard deviation, $\sigma>0$. Let $P\sim \mathcal{N}(\mu_1,\sigma)$ and $Q\sim\mathcal{N}(\mu_2,\sigma)$ be two normal distributions. Geometrically speaking, increasing $\sigma$ reduces the distance between $P$ and $Q$; this corresponds to a contraction of the space. Conversely, decreasing the variance dilates the space; as $\sigma \to 0$, the Fisher information, $\textrm{diag}(1/\sigma^2,1/\sigma^2)$, is degenerate and the distance between $P$ and $Q$ tends to infinity. \cb

Equipped with the Fisher information, we may begin to explain some foundational ideas from information geometry; including geodesic curves, geodesic distances between distributions for statistical models, and scalar curvature \cite{Amari1985}. We denote the elements of the Fisher information as $\mathcal{I}(\bm\theta) = [g_{ij}(\bm\theta)]$, and its inverse $\mathcal{I}(\bm\theta)^{-1} = [g^{ij}(\bm\theta)]$, where $\bm\theta = (\theta_1, \theta_2,...,\theta_\nu)$ are the coordinates of the manifold. While uncertainty in estimates is typically characterised by the Fisher information at only a single point, based on the Cramer-Rao inequality \cite{Watanabe2009}, information geometry utilises the Fisher information throughout the parameter space. A Riemann geodesic is a curve forming the shortest path between two points in a Riemannian manifold \cite{Amari2010}. The length of this shortest curve is referred to as the Fisher or Fisher-Rao distance \cite{Pinele2019}. We soon discuss a relationship between confidence regions and the length of geodesic curves. Informally, with greater information supporting a MLE, coinciding with an increase in its relative likelihood, confidence regions tighten. This also corresponds to a dilation of the parameter manifold; thereby increasing the geodesic distance between the MLE and other parameter combinations, reflecting their relatively reduced likelihood.     

A curve $\mathbf{z}(s)$, parameterised by $s$, connecting the points $\mathbf{z}_1 = \mathbf{z}(s_1)$ and $\mathbf{z}_2 = \mathbf{z}(s_2)$ on a Riemannian manifold, has length  \cite{Jost2017}:
\begin{align}
L(\mathbf{z}) = \int_{s_1}^{s_2} \sqrt{\sum_{i,j=1}^n \left(g_{ij}(\bm\theta(\mathbf{z}(s)))\frac{\textrm{d}\theta_i(\mathbf{z}(s))}{\textrm{d}s}\frac{\textrm{d}\theta_j(\mathbf{z}(s))}{\textrm{d}s}\right)}\textrm{ d}s. \label{eq:CurveLength}
\end{align}

A Riemann geodesic is a curve that minimises $L(\mathbf{z})$ (\ref{eq:CurveLength}), such that the distance between two points on a Riemannian manifold is given by the curve that satisfies 
\begin{align*} 
d(\mathbf{z}_1,\mathbf{z}_2)= \min \{ L(\mathbf{z}) : \mathbf{z}(s_1) = \mathbf{z}_1, \mathbf{z}(s_2) = \mathbf{z}_2 \}. 
\end{align*}

For Gaussian likelihoods, there is an \cbl asymptotic \cb  relationship between the geodesic distance between the MLE, $\boldsymbol{\hat{\theta}}$, and a point  $\boldsymbol{\theta}_\alpha$ that corresponds to an $\alpha$-level confidence region on the manifold \cite{Arutjunjan2020}. The geodesic distance between $\boldsymbol{\hat{\theta}}$ and $\boldsymbol{\theta}_\alpha$: $d(\boldsymbol{\hat{\theta}},\boldsymbol{\theta}_\alpha)$; can be written in terms of the $\alpha$th-quantile of the $\chi^2(\nu)$ distribution
\begin{align}
d(\boldsymbol{\hat{\theta}},\boldsymbol{\theta}_\alpha) = \sqrt{\Delta_{\nu,\alpha}}. \label{eq:geodesicCR}
\end{align} 

Pairing Equations (\ref{eq:LikelihoodConfRegion}) and (\ref{eq:geodesicCR}) yields an \cbl asymptotic \cb relationship between confidence regions and geodesic length \cite{Arutjunjan2020}:
\begin{align}
2\left(\ell(\boldsymbol{\hat{\theta}})-\ell(\boldsymbol{\theta}) \right) \sim d(\boldsymbol{\hat{\theta}},\boldsymbol{\theta}_\alpha)^2\quad \quad \textrm{as } N \to \infty.\label{eq:geoCRdist}
\end{align}

In Section \ref{sec:Results} we present likelihood-based confidence regions alongside geodesic curves of the corresponding length, as characterised by (\ref{eq:geodesicCR}), and comment on the validity of Equation (\ref{eq:geoCRdist}) in a range of scenarios.  

Geodesic curves satisfy the following system of differential equations in $n$ dimensions \cite{Pinele2020}:
\begin{align}
\frac{\textrm{d}^2\theta_m}{\textrm{d}s^2} &+ \sum_{i,j=1}^n\Gamma^m_{ij}\frac{\textrm{d}\theta_i}{\textrm{d}s}\frac{\textrm{d}\theta_j}{\textrm{d}s} = 0, \quad m = 1,...,n,\label{eq:GeodesicODE}   
\end{align}
where $s$ is the parameterisation of the geodesic curve, in accordance with Equation (\ref{eq:CurveLength}), and $\Gamma^m_{ij}$ are the Christoffel symbols of the second kind \cite{Calin2014}, defined as 
\begin{align}
\Gamma^m_{ij} &= \dfrac{1}{2} \sum_{l=1}^ng^{ml}\left(\frac{\partial g_{lj}}{\partial \theta_i}+\frac{\partial g_{li}}{\partial \theta_j}-\frac{\partial g_{ij}}{\partial \theta_l}\right). \label{eq:Christoffel}
\end{align}

We can convert from Christoffel symbols of the second kind to Christoffel symbols of the first kind by lowering the contravariant (upper) index through multiplication by the metric: $\Gamma_{kij} = g_{km}\Gamma^m_{ij}$\cite{Heinbockel2001}. Here, repeated indices, in this case $m$, imply a summation is to be performed over the repeated index, following the Einstein summation convention \cite{Amari2016}. Conversely, we can recover Christoffel symbols of the first kind from Christoffel symbols of the second kind via the inverse metric:  $g^{km}\Gamma_{kij} = \Gamma^m_{ij}$. Christoffel symbols of the second kind are the connection coefficients of the Levi-Civita connection; the Christoffel symbols are symmetric in the covariant (lower) indices \cite{Giesel2021}. On an $n$-dimensional manifold, the Christoffel symbol is of dimension $n \times n \times n$. Geodesics can be used to construct theoretical confidence regions, to measure the geometric distance between probability distributions, and to perform hypothesis testing, for example to test equality of parameters \cite{Kass1997,Menendez1995,Nielsen2020}.  

Under certain conditions, analytical expressions can be obtained for the solutions of the geodesic equations, and the corresponding Fisher-Rao distances, for example in the case of the univariate (\ref{eq:UnivNormal}) and multivariate (\ref{eq:MvNormal}) normal distributions \cite{Calvo1991,Pinele2019}. However, we solve Equation (\ref{eq:GeodesicODE}) numerically, after converting the second order ODE to a first order system of ODEs using standard techniques. 

We are also interested in exploring the scalar curvature, also known as the \textit{Ricci scalar}, of our manifolds. To compute the scalar curvature, we must first construct the Riemann tensor, and subsequently the Ricci tensor. As we only require these tensors for computation of the scalar curvature, and do not attempt to interpret these tensors directly in this work, we provide only a limited outline of their interpretation.  The Riemann curvature tensor is constructed from the Christoffel symbols and their first partial derivatives. Here, it is convenient to think about these partial derivatives as being with respect to the parameters of interest. Due to the possibility of raising or lowering indices of Christoffel symbols and tensors via the metric, there are several equivalent expressions for computing the Riemann curvature tensor \cite{Heinbockel2001}. The elements of the Riemann tensor of the first kind can be written as 
\begin{align} 
R_{ijkl}= \frac{\partial \Gamma_{jli}}{\partial k} - \frac{\partial \Gamma_{jki}}{\partial l} + \Gamma_{ilr}\Gamma^r_{jk} - \Gamma_{ikr}\Gamma^r_{jl}.\label{eq:RiemannTensor}
\end{align}
The Riemann tensor of the first kind is a (0,4) tensor (with no contravariant indices and four covariant indices), and can be converted to the (1,3) Riemann tensor of the second kind via the inverse of the metric: $g^{im}R_{ijkl} = R^m_{jkl}$. On an $n$-dimensional manifold, the Riemann tensor is of dimension $n \times n \times n \times n$; due to various symmetries however, there are far fewer independent elements \cite{Luscombe2018}. The Riemann tensor provides information about the intrinsic curvature of the manifold. A geometric interpretation is that a vector from a point on the manifold, parallel transported around a parallelogram, will be identical to its original value when it returns to its starting point if the manifold is flat. In this case, the Riemann tensor vanishes. If the manifold is not flat, the Riemann tensor can be used to quantify how the vector differs following this parallel transport \cite{Loveridge2016}.   

From the Riemann tensor of the second kind, we can compute the Ricci tensor of the first kind. The Ricci tensor, $R_{ij}$, is obtained by contracting the contravariant index with the third covariant index of the Riemann tensor of the second kind; that is,
\begin{align}
R_{ij} = R^m_{ijm}.\label{eq:RicciTensor}
\end{align} 
On an $n$-dimensional manifold, the Ricci tensor is of dimension $n \times n$, and is symmetric \cite{Loveridge2016}. The Ricci tensor can quantify the changes to a volume element as it moves through the manifold, relative to Euclidean space \cite{Loveridge2016}. 

The scalar curvature, $Sc$, can be obtained as a contraction of the Ricci tensor
\begin{align} 
	Sc = g^{ij}R_{ij}.\label{eq:ScalarCurvature}
\end{align}
The scalar curvature is invariant; it does not change under a change of coordinates (re-parameterisation). \cbl For Gaussian likelihoods, the corresponding manifold is flat; characterised by zero scalar curvature everywhere. \cb As such, the scalar curvature provides a measure of how the likelihood of the underlying statistical model deviates from being Gaussian---often referred to as \textit{non-Gaussianity} in the physics and cosmology literature---irrespective of the parameterisation \cite{Giesel2021}. As we will explore in Section \ref{sec:Results}, it can also provide insights into parameter identifiability.   

\subsection{Hypothesis testing}

Here we outline the approach for performing likelihood-ratio-based hypothesis tests, and hypothesis tests based on geodesic distance. As we consider synthetic data in this work, we know the \textit{true} parameter values, $\boldsymbol{\theta}_\textrm{t}$. In practical applications this is not the case. As such, we may seek to test whether some previously held notion about the true parameters, $\boldsymbol{\theta}_\textrm{t}= \boldsymbol{\theta}_0$, is supported by the data, based on the computed MLE. This could be investigated via the following hypothesis test: 
\begin{align}
\begin{cases}
H_0: \boldsymbol{\theta}_\textrm{t} = \boldsymbol{\theta}_0,\\
H_1: \boldsymbol{\theta}_\textrm{t} \ne \boldsymbol{\theta}_0.
\end{cases}
\end{align}
From Equation (\ref{eq:LikelihoodConfRegion}) the test statistic for such a likelihood-rato-based hypothesis test can be expressed as 
 \begin{align}
 	\lambda_{LR} = -2(\ell(\boldsymbol{\theta}_0)-\ell(\hat{\boldsymbol{\theta}})), \label{eq:LLhoodTest}
 \end{align}
where asymptotically as $N \rightarrow \infty$, $\lambda_{LR}\sim \chi^2(\nu)$,  following Wilk's theorem \cite{Pawitan2001}. From the asymptotic relationship given in Equation (\ref{eq:geoCRdist}), it follows that under the same asymptotic relationship the test statistic for a hypothesis test based on geodesic distance is \cite{Kass1997}:
 \begin{align}
\lambda_{GD} = d(\boldsymbol{\theta}_0,\hat{\boldsymbol{\theta}})^2. \label{eq:GeodesicTest}
\end{align}

The likelihood values required to compute Equation (\ref{eq:LLhoodTest}) can be obtained directly by evaluating Equation (\ref{eq:llfun}). To compute the geodesic distance between two specific points in parameter space, as required by Equation (\ref{eq:GeodesicTest}), it is necessary to solve a boundary value problem to obtain the geodesic curve between $\boldsymbol{\theta}_0$ and $\hat{\boldsymbol{\theta}}$. Approximate p-values can be computed from these test statistics as $1-F_{\chi^2(\nu)}(\lambda_{LR})$ and $1-F_{\chi^2(\nu)}(\lambda_{GD})$, respectively,  where  $F_{\chi^2(\nu)}$ is the cumulative distribution function of $\chi^2(\nu)$ \cite{Browning2021a}. We provide practical examples of each of these approaches to hypothesis testing in Section \ref{sec:Results}.

\subsection{Numerical implementation}

All numerical techniques used to produce the results in this work are implemented in the open source Julia language \cite{Bezanson2012}; we use a combination of existing Julia packages and bespoke implementations. There are several aspects of numerical computation in this work, including approximate solutions to systems of ODEs, differentiation with both finite differences and forward mode automatic differentiation, likelihood computation and nonlinear optimisation. Nonlinear optimisation for obtaining MLEs and parameter combinations corresponding to particular confidence levels is performed with the Julia package NLopt.jl, using the Bound Optimisation by Quadratic Approximation (BOBYQA) algorithm. BOBYQA is a derivative-free algorithm for solving bound constrained optimisation problems \cite{Powell2009}. Approximate solutions to ODEs are obtained using the Julia package DifferentialEquations.jl \cite{Rackauckas2017}. The second order Heun's method \cite{Dobrushkin2017}; a two-stage Runge-Kutta method, is used for obtaining contours of the log-likelihood function to form approximate likelihood-based confidence regions \cite{Browning2021a}. Heun's method is implemented as \texttt{Heun()} in DifferentialEquations.jl. Approximate solutions to geodesic differential equations are obtained using the Tsitouras implementation of the Runge–Kutta method, that employs Runge–Kutta pairs of orders 5 and 4 \cite{Tsitouras2011}, implemented as \texttt{Tsit5()} in DifferentialEquations.jl. Boundary value problems for geodesic-distance-based hypothesis tests are solved using the DifferentialEquations.jl implementation of a shooting method, utilising \texttt{Tsit5()}. Code for reproducing all examples in this work is available on \href{https://github.com/Jesse-Sharp/Sharp2021b}{GitHub}.

\section{Results} \label{sec:Results}

In this section we present results combining likelihood based parameter inference and uncertainty quantification with ideas from information geometry, including geodesic curves and scalar curvature. We apply these techniques to univariate and multivariate normal distributions, linear, exponential and logistic population growth models and the SIR model. Through these canonical examples, we explore pedagogically differences in the inference and information geometry results that arise as we consider parameter estimation and uncertainty for increasingly complex systems. 

Synthetic data for the univariate and multivariate normal distributions are generated by sampling from the respective distributions given in Equation (\ref{eq:NormalDists}). For simplicity, in this work we consider synthetic data from uncorrelated observation processes with constant standard deviation in both time and parameter space. However, we note that the techniques presented in this work can be generalised to handle data with non-constant variance and for other distributions where the Fisher information is available \cite{Browning2021b}.       
\begin{align}
\textrm{Univariate}: x_i \sim \mathcal{N}(\mu,\sigma^2), \quad \textrm{Multivariate}: \mathbf{x}_i \sim \textrm{MVN}(\boldsymbol{\mu},\boldsymbol{\Sigma}),\label{eq:NormalDists}
\end{align}
where $\boldsymbol{\Sigma} = \textrm{diag}(\sigma^2)$ is the covariance matrix. For the population growth and SIR models considered in this work, synthetic data is generated by drawing from a normal distribution with mean described by the model solution and a prescribed standard deviation; effectively substituting $\mu = \mu(\boldsymbol{\theta},t)$ in Equation (\ref{eq:NormalDists}) for observation processes with a single variable, and $\boldsymbol{\mu} = \boldsymbol{\mu}(\boldsymbol{\theta},t)$ for observation processes with several variables. When $\sigma$ is one of the parameters to be estimated, $\sigma \in \boldsymbol{\theta}$, but $\boldsymbol{\mu}$ does not depend on $\sigma$. Parameter values that we use to generate synthetic data correspond to parameter estimates inferred from field data in the literature \cite{Murray2002,Simpson2021}. 

We present a series of figures in this section visualising the normalised log-likelihood, $\hat{\ell}$, and scalar curvature, $Sc$, as heatmaps, with likelihood-based 95\% confidence regions and geodesics with a length corresponding to a 95\% confidence distance superimposed. All results are computed numerically, as outlined in Section \ref{sec:Methods}, with code available on \href{https://github.com/Jesse-Sharp/Sharp2021b}{GitHub}. Unless otherwise indicated, each set of geodesics includes 20 geodesics with initial velocities corresponding to equidistant points uniformly distributed on the circumference of a unit circle. As such, the apparent clustering of geodesics in some examples highlights differences in the scaling and stretching of parameter spaces. Each scalar curvature and log-likelihood heatmap is computed on a uniformly discretised $100\times100$ grid.

\subsection{Normal distributions}

\begin{figure}
	\centering
	\includegraphics[width=\linewidth]{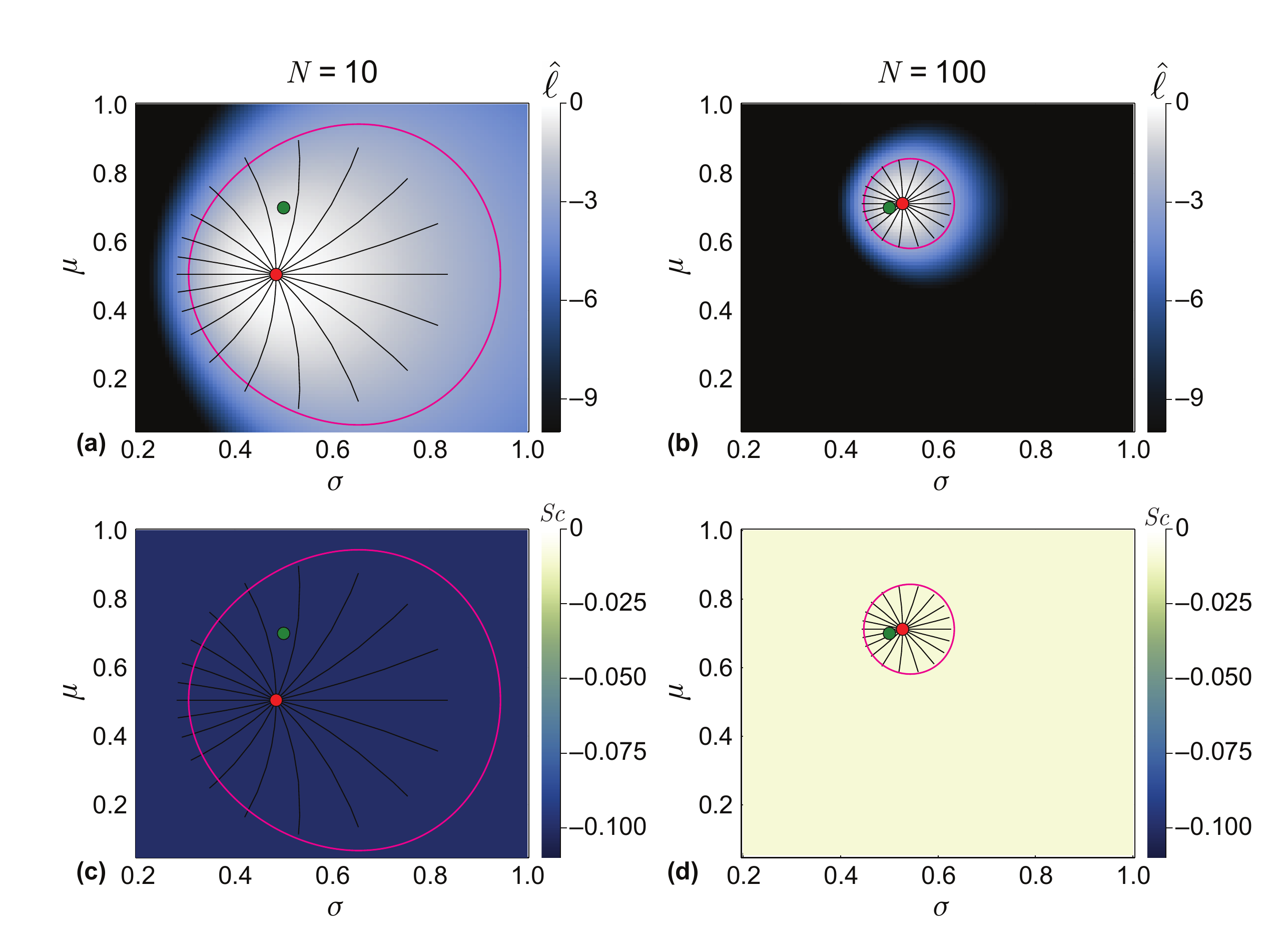}
	\caption{Univariate normal distribution with inferred mean, $\mu$, and standard deviation, $\sigma$. Heatmaps visualise the normalised log-likelihood, $\hat{\ell}$, (a,b) and the scalar curvature, $Sc$, (c,d). True parameter values, $(\mu,\sigma) = (0.7,0.5)$, are marked with green discs, with the MLEs indicated using red discs. Magenta curves correspond to likelihood-based 95\% confidence regions. Black lines are geodesic curves emanating from the MLEs, with a geodesic length corresponding to a theoretical 95\% confidence region. Increasing the number of data points, $N$, tightens the confidence regions, improves the correspondence between geodesic curves and likelihood-based confidence regions, and reduces the scalar curvature. \label{fig:UnivariateNormal}}
\end{figure}

We first consider parameter inference and information geometry techniques applied to observations drawn directly from univariate and bivariate normal distributions, with no underlying process model. In Figure \ref{fig:UnivariateNormal} we present results for the univariate normal distribution (\ref{eq:UnivNormal}), estimating $\boldsymbol{\theta} = (\mu,\sigma)$. The true mean and standard deviation used to generate data are $(\mu,\sigma) = (0.7,0.5)$. Estimates are obtained via maximum likelihood estimation. MLEs of normal variance are known to provide biased underestimates \cite{Pawitan2001}, and the derivation of the Fisher information assumes an unbiased estimator \cite{Frieden2007}. This may partially explain the particular differences observed between the likelihood-based confidence region and the endpoints of the geodesics in Figure \ref{fig:UnivariateNormal}, wherein the geodesics appear not only to suggest a tighter confidence region, but also appear to be biased towards parameter space with smaller standard deviation. As the number of observations increases from $N=10$ to $N = 100$, we observe not only that the MLE more precisely estimates the true parameter values, but also the endpoints of the geodesic curves more closely correspond to the likelihood-based confidence regions. This is consistent with both the theoretical asymptotic relationship between geodesic length and likelihood-based confidence regions given in Equation (\ref{eq:geoCRdist}), and also the bias of the MLE for standard deviation decreasing, as $N$ increases. 

The manifold representing the family of normal distributions parameterised by $\boldsymbol{\theta} = (\mu,\sigma)$ has constant scalar curvature $Sc = -1$. Due to the additive nature of the Fisher information, having $N$ observations results in a constant scalar curvature of $Sc = -1/N$, as presented in Figure \ref{fig:UnivariateNormal}(c,d). It is straightforward, although tedious, to verify this result through combining Equations (\ref{eq:FIM}), (\ref{eq:Christoffel}), (\ref{eq:RiemannTensor}), (\ref{eq:RicciTensor}) and (\ref{eq:ScalarCurvature}).     

\FloatBarrier

\begin{figure}[h]
	\centering
	\includegraphics[width=\linewidth]{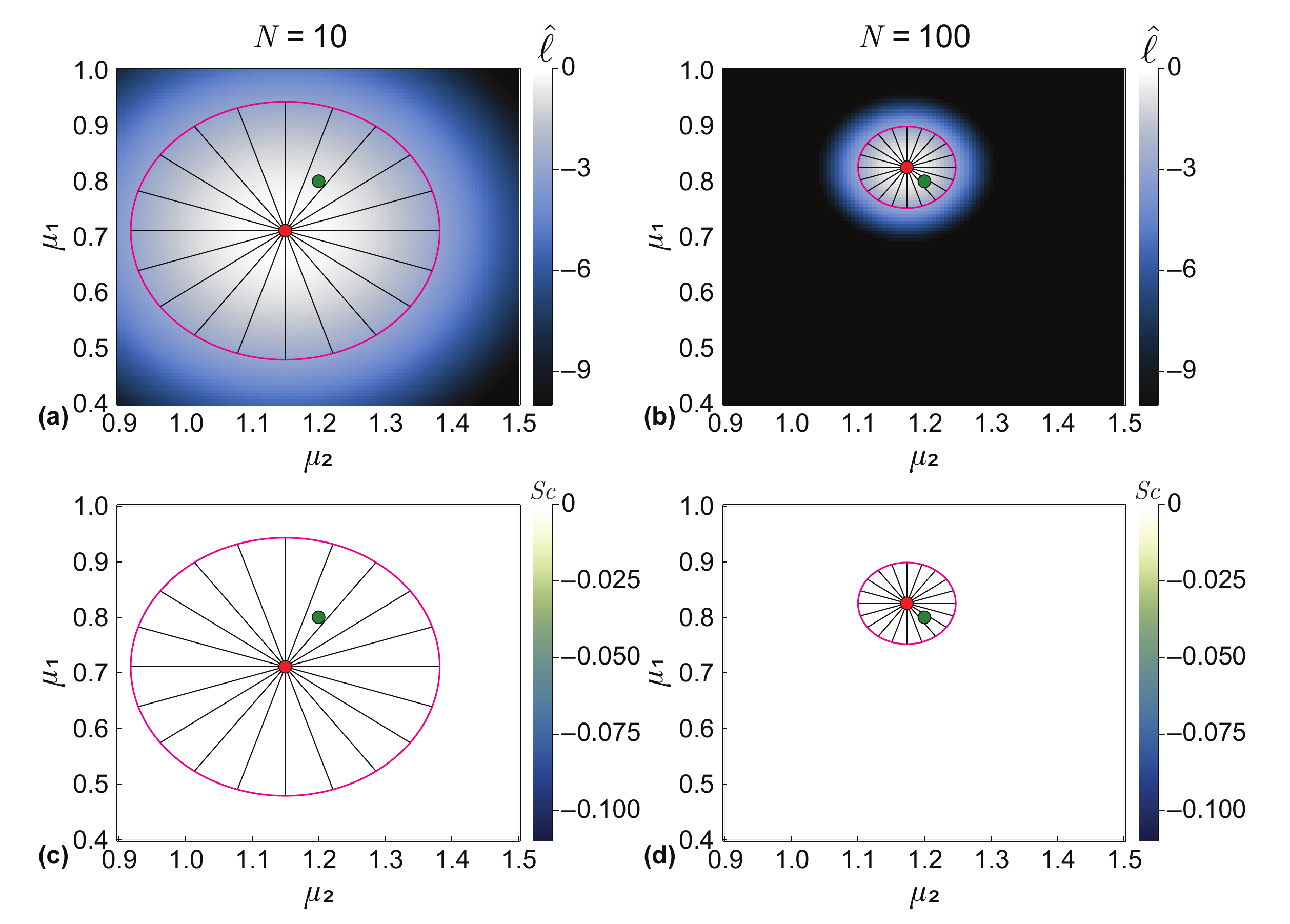}
	\caption{Multivariate normal distribution with inferred means, $\mu_1$, and $\mu_2$, with known constant standard deviation, $\sigma = 0.3$. Heatmaps visualise the normalised log-likelihood, $\hat{\ell}$, (a,b) and the scalar curvature, $Sc$, (c,d). True parameter values, ($\mu_1,\mu_2)=(0.8,1.2)$, are marked with green discs, with the MLEs indicated using red discs. Magenta curves correspond to likelihood-based 95\% confidence regions. Black lines are geodesic curves emanating from the MLEs, with geodesic lengths corresponding to a theoretical 95\% confidence region. Increasing the number of data points, $N$, tightens the confidence regions. In contrast to the univariate case where we infer standard deviation in Figure \ref{fig:UnivariateNormal}, when only inferring the mean parameters of the multivariate normal distribution, we see that even with few observations, $N = 10$, the geodesics and likelihood-based confidence regions match closely. As we are estimating means only, and there is no model-induced curvature, the scalar curvature is zero everywhere. \label{fig:MvNormal} }
\end{figure}

The probability density function for the multivariate normal distribution with two independent variables; $x,y \in \mathbb{R}$, with constant standard deviation $\sigma$ is  
\begin{align}
p(x,y;\mu_1,\mu_2,\sigma) = \frac{1}{2\pi \sigma^2}\exp\left({-\left(\frac{(x-\mu_1)^2+(y-\mu_2)^2}{2\sigma^2}\right)}\right). \label{eq:MvNormal}
\end{align}     
In Equation (\ref{eq:MvNormal}) there are 3 parameters that we could estimate from data; $\boldsymbol{\Theta} = (\mu_1,\mu_2,\sigma)$. As we estimated the mean and standard deviation for the univariate normal distribution in Figure \ref{fig:UnivariateNormal}, we consider inference of both means for the multivariate normal; $\boldsymbol{\theta} = (\mu_1,\mu_2)$. Results are presented in Figure \ref{fig:MvNormal}. Even with a small number of observations ($N = 10$), we observe an excellent match between the likelihood-based confidence regions and geodesics when only estimating means. As expected, increasing $N$ results in a MLE closer to the true values, and tighter confidence regions. We also observe that the confidence regions are symmetric with respect to each mean parameter. When estimating only the mean parameters of the multivariate normal distribution, we see that the scalar curvature is zero everywhere. This is to be expected, as the Fisher information for normally distributed observation processes, Equation (\ref{eq:normalFIM}), depends only on the standard deviation and not the mean. As such all of the partial derivatives used to construct the Christoffel symbols, (\ref{eq:Christoffel}), are zero; this vanishing of the Christoffel symbols translates to zero scalar curvature through Equations (\ref{eq:RiemannTensor}), (\ref{eq:RicciTensor}) and (\ref{eq:ScalarCurvature}). We also observe that, in contrast to the evident curvature of the geodesics for the univariate normal case presented in Figure \ref{fig:UnivariateNormal}, the geodesic curves in Figure \ref{fig:MvNormal} appear perfectly straight when plotted in Euclidean geometry. The Riemann tensor (\ref{eq:RiemannTensor}) is zero everywhere when inferring multivariate normal means. This suggests that the manifold is flat.   

\begin{figure}[h]
	\centering
	\includegraphics[width=0.9\linewidth]{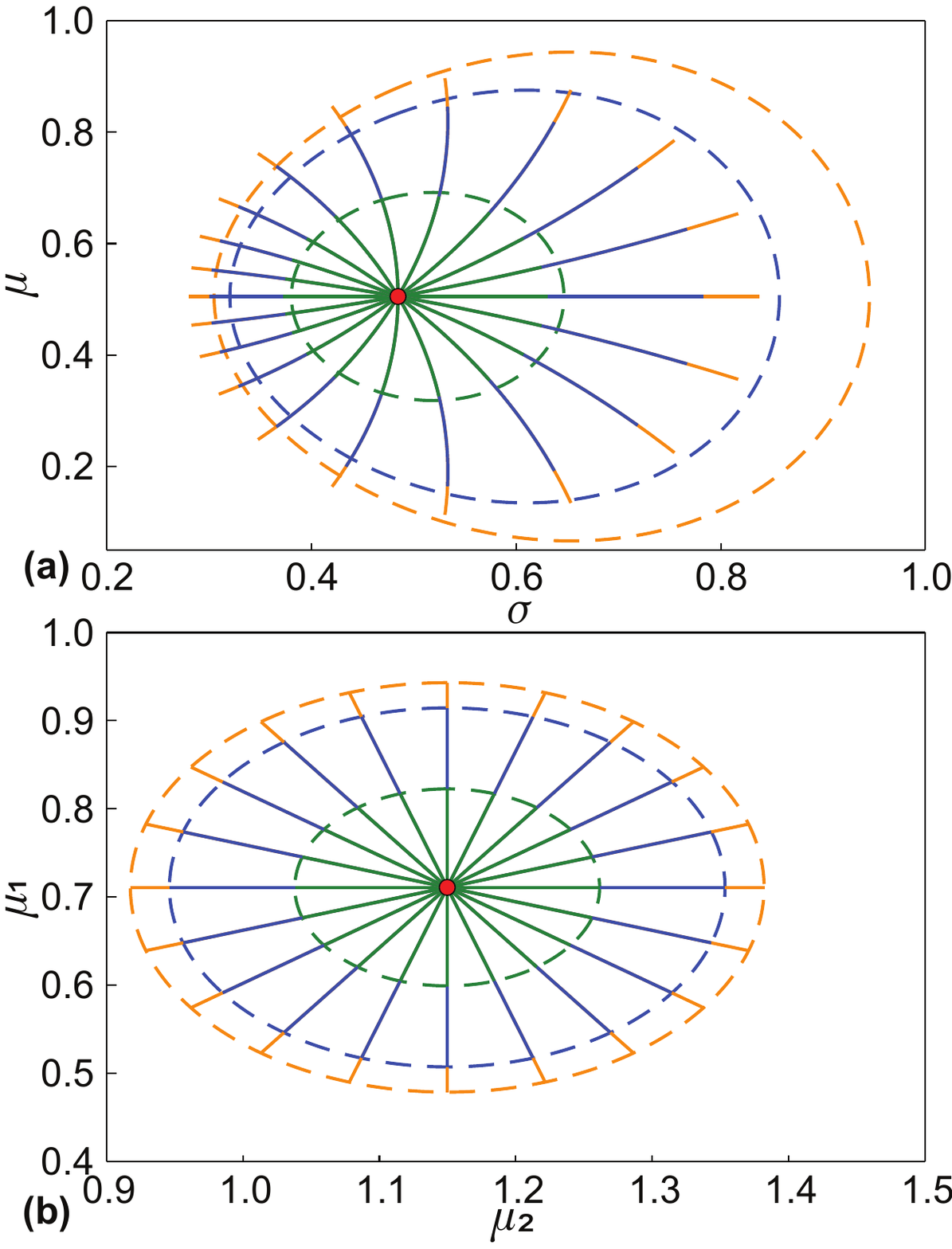}
	\caption{\cbl Comparison of confidence regions at intermediate likelihood values and geodesic distances. Results correspond to (a) univariate normal distribution with inferred mean, $\mu$, and standard deviation, $\sigma$, as considered in Figure \ref{fig:UnivariateNormal}a, and (b) multivariate normal distribution with inferred means, $\mu_1$, and $\mu_2$, as considered in Figure \ref{fig:MvNormal}a. MLEs are indicated using red discs. Dashed curves correspond to likelihood-based 50\% (green), 90\% (blue), and 95\% (orange) confidence regions. Solid lines are geodesic curves emanating from the MLEs, with geodesic lengths within a theoretical 50\% (green), 90\% (blue), and 95\% (orange) confidence distance. \cb \label{fig:NormalModelCRs} }
\end{figure}

\cbl Results presented in this work predominantly feature 95\% confidence regions. We note that although this choice is common \cite{Fisher1992}, it is also arbitrary, and equivalent analysis could be performed at different confidence levels. In examples where the geodesic endpoints approximately align with the likelihood-based confidence regions at the 95\% level, we expect intermediate points along the geodesics to also approximately align with corresponding likelihood contours, in accordance with Equation (\ref{eq:geoCRdist}). However, in examples where we observe a mismatch between geodesic endpoints and likelihood-based confidence intervals at the 95\% level, we do not expect intermediate points along geodesics to correspond to likelihood contours. This is demonstrated in Figure \ref{fig:NormalModelCRs}.    \cb

Having considered the techniques as applied directly to distributions, we now incorporate ODE-based process models, such that our observations are normally distributed about the solution of a mathematical model.   

\FloatBarrier
\clearpage
\subsection{Population growth models}

\begin{figure}[h]
	\centering
	\includegraphics[width=0.9\linewidth]{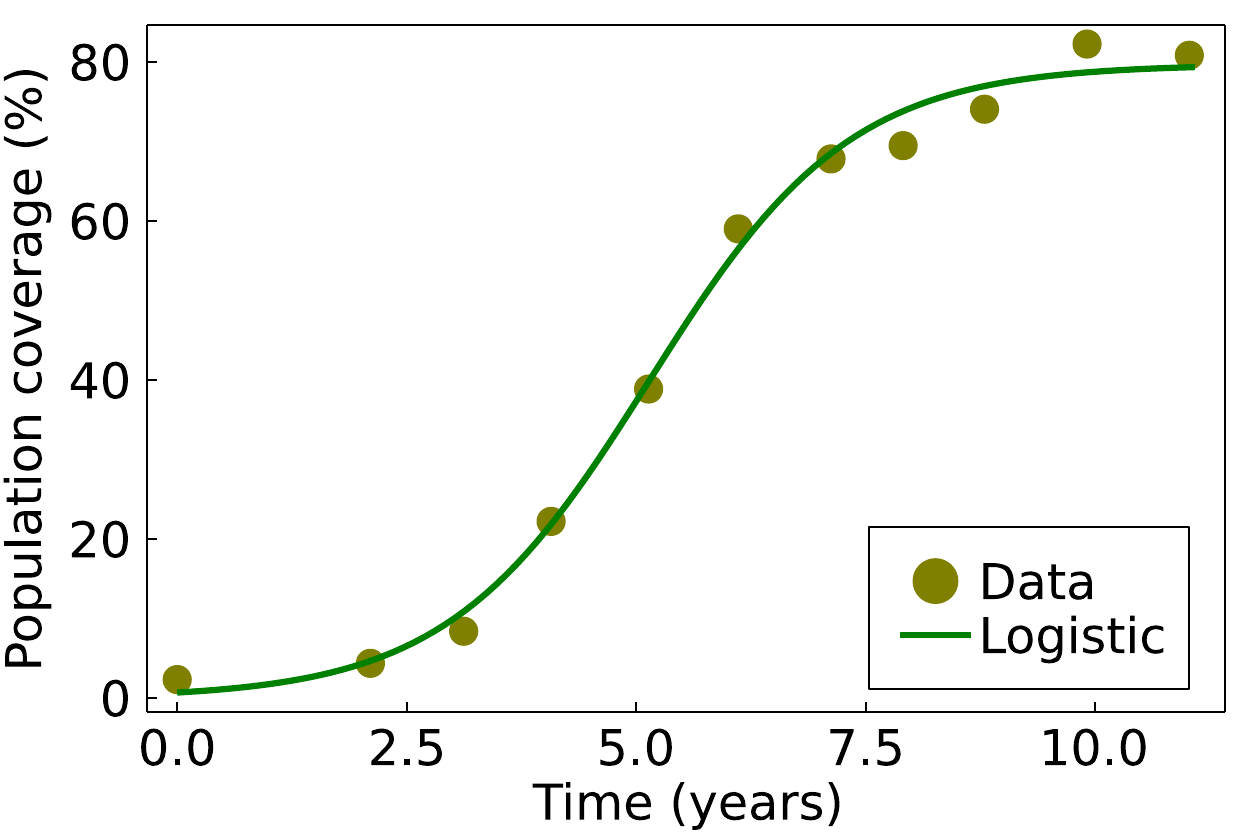}
	\caption{Markers correspond to data from field studies, representing the \% of area in a region covered by hard corals, as the coral population regrows following depletion by an external event \cite{Simpson2021}. Data originally extracted from the Australian Institute of Marine Science (AIMS) Long Term Monitoring Program (LTMP) eAtlas (eatlas.org.au/gbr/ltmp-data). A logistic model is fit to the data in \cite{Simpson2021}, with inferred parameters: $r = 0.9131$ [year$^{-1}$], $C(0) = 0.7237\%$, $K = 79.74\%$ and standard deviation $\sigma =2.301$; this is reproduced here as the green curve. \label{fig:LogisticData} }
\end{figure}

The canonical logistic growth model, alongside generalisations and related sigmoid models such as Gompertz and Richards' models, have been extensively applied to study population growth dynamics in the life sciences \cite{Panik2014,Simpson2021}. In Figure \ref{fig:LogisticData} we present data from the literature describing the area covered by hard corals in a region as they regrow following an adverse event. This can be modelled as a logistic growth process \cite{Simpson2021}. Logistic growth of a population with density, $C(t)$, is characterised by a growth rate, $r>0$, initial condition, $C(0)>0$, and carrying capacity, $K>0$. Treating parameters values ($r$,$C(0)$,$K$) =  $(0.9131$ [year$^{-1}], 0.7237\%, 79.74\%)$, and standard deviation $\sigma = 2.301\%$, inferred in the literature from this field data as the \textit{true} values, we generate various synthetic data sets with multiple observations at various time-points.

\begin{figure}[H]
	\centering
	\includegraphics[width=0.6\linewidth]{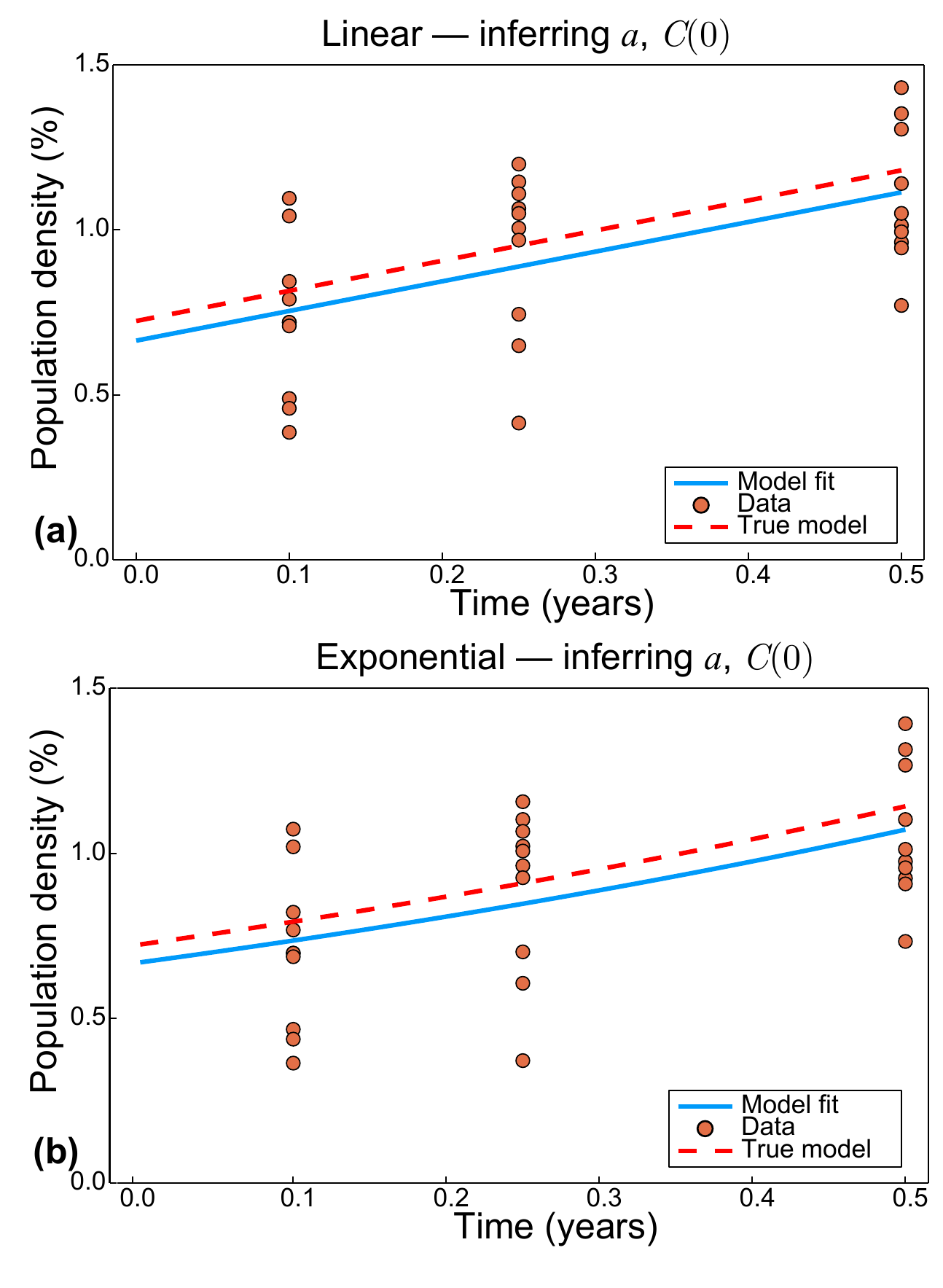}
	\caption{Example synthetic data generated from the linear and exponential models with comparison of early time linear and exponential model fits, inferring $a$ and $C(0)$. $N = 10$ observations per time-point, with time-points $T = (0.1,0.25,0.5)$. True parameter values are $a = 0.9131$, $C(0) = 0.7237$, with known standard deviation, $\sigma = 0.2301$. For generating synthetic early time linear and exponential data, we reduce the standard deviation relative to the $\sigma = 2.301$ computed from the logistic model, as early time data produced with $C(0) = 0.7237$ and $\sigma = 2.301$ produces negative population density observations. Inference produces MLEs of $(\hat{a},\hat{C(0)}) = (0.8988,0.6642)$ for the linear model, and $(\hat{a},\hat{C(0)}) = (0.9412,0.6695)$ for the exponential model.   \label{fig:ExpLinFit}    }
\end{figure}

\begin{figure}[H]
	\centering
	\includegraphics[width=0.7\linewidth]{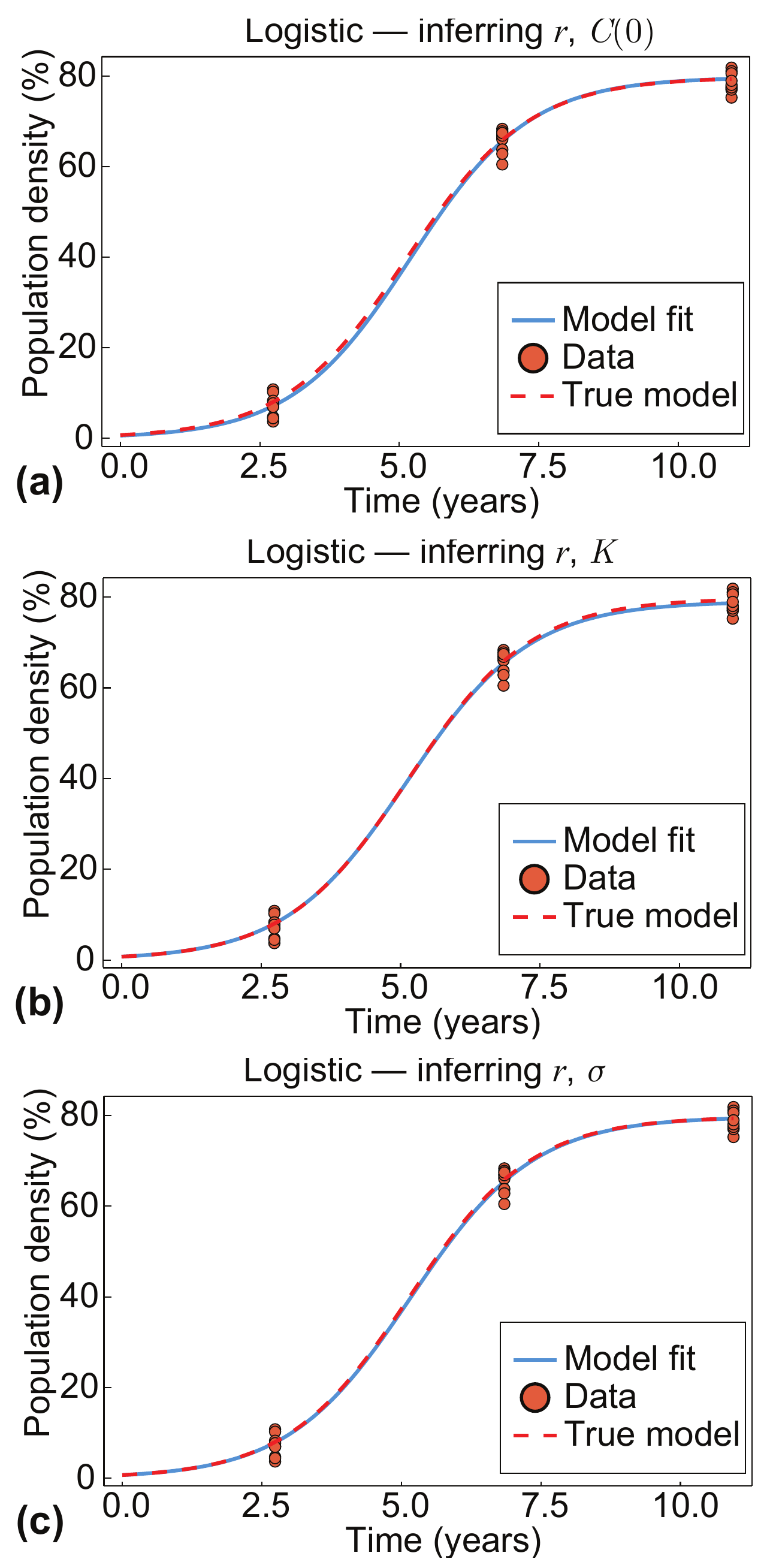}
	\caption{Example synthetic data generated from the logistic growth model. The logistic model is fit to the synthetic data, inferring pairwise combinations of $r$ with $C(0)$, $K$ and $\sigma$. Observations are made at $T = (2.74,6.84,10.95)$ years, with $N = 10$ observations per time-point. True parameter values are $r = 0.9131$, $C(0) = 0.7237$, $K = 79.74$ and $\sigma = 2.301$. \label{fig:LogisticFit}}
\end{figure}

The logistic growth model is well approximated by the exponential growth model when $C(t)\ll K$ \cite{Tsoularis2002}, and early time exponential growth is approximately linear. Before considering the inference and information geometry techniques as applied to the logistic model, we first consider the more fundamental linear and exponential growth models. In Figure \ref{fig:ExpLinFit} we present example synthetic linear and exponential data, and in Figure \ref{fig:LogisticFit} synthetic logistic data.  In the context of population growth models, the presence of variability in observations at a single time-point could reflect, for example; measurement error, variability in population estimates, or expert judgement \cite{Ferson1996}.

\clearpage

\subsubsection{Linear growth}

Linear growth describes growth at a constant rate, independent of the population density. The linear growth model and solution are given by 

\begin{align*}
\frac{\textrm{d}C}{\textrm{d}t} &= a, \quad C(t) = at+C(0).
\end{align*}

With parameters $\boldsymbol{\Theta} = (a,C(0),\sigma)$, $\mu(\boldsymbol{\Theta},t) =  at+C(0)$ describes the expected model behaviour. In Figure \ref{fig:LinExp12panel}(a-f) we present inference results for the linear model for all pairwise combinations of $\boldsymbol{\Theta}$. The partial derivatives of the linear model with respect to the parameters $a$ and $C(0)$, required to form the Jacobians, $\mathbf{J}(\boldsymbol{\theta})$, are 

\begin{align*} 
\frac{\partial \mu(\boldsymbol{\Theta},t)}{\partial a} & = t, \quad
\frac{\partial \mu(\boldsymbol{\Theta},t)}{\partial C(0)}  = 1.	
\end{align*} 

Recall from Equation (\ref{eq:modelJAC}) that we only require the partial derivatives corresponding to unknown parameters in any given example. When estimating $\boldsymbol{\theta}$ = $(a,C(0))$ we find that, similar to the multivariate normal case where we estimate means, the scalar curvature is zero everywhere. We also observe that the end-points of the geodesics align with the likelihood-based confidence region. We stress that this arises through the relationship in Equation (\ref{eq:geoCRdist}), and is not forced to occur via termination of the numerical solution of the ODE once it reaches the likelihood-based confidence region. However, due to the relationship between $a$ and $C(0)$, we find that the confidence regions in this case are not symmetric about the MLE with respect to each parameter. Rather, we see that for a given normalised log-likelihood value a larger growth rate corresponds to a smaller initial condition, and vice versa. This aligns with our intuition when considering fitting a straight line through data, as presented in Figure \ref{fig:ExpLinFit}a; lines with a greater slope ($a$) must start lower ($C(0)$) to fit the data.

When one of the parameters to be estimated is $\sigma$, we observe similar results to the univariate normal case; geodesic endpoints are offset in the direction of decreasing $\sigma$ relative to the likelihood-based confidence regions, and there is constant scalar curvature of $Sc=-1/N$. The geodesics and confidence regions appear symmetric with respect to the model parameter, about the MLE. 

\subsubsection{Exponential growth}
Exponential growth describes growth at a rate proportional to the size of the population. The exponential growth model and solution are

\begin{align*}
\frac{\textrm{d}C}{\textrm{d}t} &= aC,\quad C(t) = C(0) \exp(at).
\end{align*}

With parameters $\boldsymbol{\Theta} = (a,C(0),\sigma)$, $\mu(\boldsymbol{\Theta},t) =  C(0) \exp(at)$ describes the expected model behaviour. The partial derivatives of the exponential model with respect to the parameters $a$ and $C(0)$, required to form the Jacobians, $\mathbf{J}(\boldsymbol{\theta})$, are 

\begin{align*} 
\frac{\partial \mu(\boldsymbol{\Theta},t)}{\partial a} & = t C(0) \exp(at), \quad 
\frac{\partial \mu(\boldsymbol{\Theta},t)}{\partial C(0)}  = \exp(at).	
\end{align*} 

\begin{figure}
	\centering
	\includegraphics[width=\linewidth]{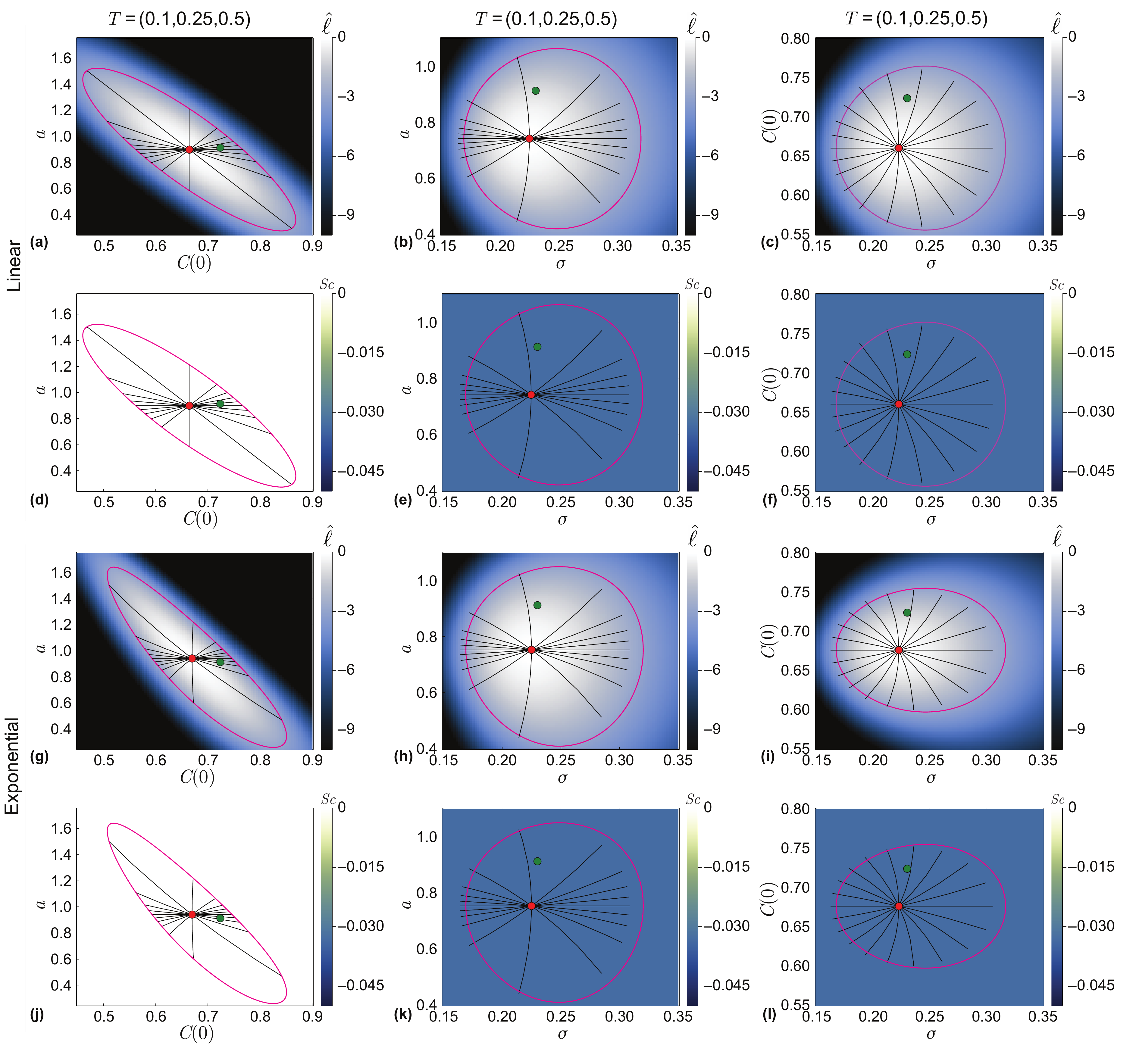}
	\caption{Linear (a-f) and exponential (g-l) models with inferred pairwise combinations of growth rate, $a$, initial condition, $C(0)$, and standard deviation, $\sigma$. Heatmaps visualise the normalised log-likelihood, $\hat{\ell}$, (a-c,g-i); and the scalar curvature, $Sc$, (d-f,j-l). Observations are made at $T = (0.1,0.25,0.5)$, with 10 observations per time-point; corresponding to the example data presented in Figure \ref{fig:ExpLinFit}.   The true parameter values are marked with green discs, with the MLEs indicated using red discs. Magenta curves correspond to likelihood-based 95\% confidence regions. Black lines are geodesic curves emanating from the MLEs, with lengths corresponding to a theoretical 95\% confidence distance. True values of model parameters correspond to the logistic growth parameters; $a = 0.9131$, $C(0) = 0.7237$, with reduced standard deviation $\sigma = 0.2301$. \label{fig:LinExp12panel}} 
\end{figure}

By construction, as detailed in Figure \ref{fig:ExpLinFit}, the linear and exponential models with identical parameters and initial conditions produce very similar behaviours over a sufficiently small time-scale. This is seen when comparing the inference results for the exponential model, presented in Figure \ref{fig:LinExp12panel}(g-l), to the corresponding linear results in Figure \ref{fig:LinExp12panel}(a-f). When inferring $\boldsymbol{\theta} = (a,\sigma)$, deviations from the corresponding linear results are minimal. The likelihood-based confidence region and corresponding geodesic endpoints for $\boldsymbol{\theta} = (a,C(0))$ are marginally tighter and less elliptical. When inferring $\boldsymbol{\theta} = (C(0),\sigma)$, we find that the confidence region for the exponential model is narrower with respect to $C(0)$ than that of the linear model, though near-identical with respect to $\sigma$. As for the linear case, the scalar curvature is $Sc = -1/N$ everywhere when $\sigma$ is one of the unknown parameters, and zero everywhere otherwise.

\FloatBarrier

\subsubsection{Logistic growth}

Logistic growth describes growth at a rate dependent on the size of the population, with growth ceasing once the population reaches a carrying capacity. For sufficiently small populations relative to the carrying capacity, logistic growth is approximately exponential \cite{Tsoularis2002}. As the population approaches the carrying capacity, the rate of growth slows. The logistic growth model is  
\begin{align*}
\frac{\textrm{d}C(t)}{\textrm{d}t} & = rC(t)\left(1-\frac{C(t)}{K}\right), 
\end{align*}
with solution
\begin{align}
C(t) &= \frac{C(0) K}{C(0) + (K-C(0))\exp\left(-rt\right)}.  \label{Eq:LogisticSol}
\end{align}

The long-time limit of Equation (\ref{Eq:LogisticSol}) is  $\lim\limits_{t\to\infty}C(t) = K$. The behaviour of the logistic model can be described by the three model parameters and standard deviation: $\boldsymbol{\Theta} = (r,C(0),K,\sigma)$. We can compute the partial derivatives required to form the Jacobian matrices, $\mathbf{J}(\boldsymbol{\theta})$, analytically:
\begin{align}
\mu(\boldsymbol{\Theta},t) &= C(r,C(0),K,t) = \frac{C(0) K}{C(0) + (K-C(0))\exp\left(-rt\right)},\nonumber\\
\frac{\partial \mu(\boldsymbol{\Theta},t)}{\partial r} &= \frac{C(0)Kt(K-C(0))\exp\left(-rt\right)}{((K-C(0))\exp\left(-rt\right)+C(0))^2},\nonumber\\
\frac{\partial \mu(\boldsymbol{\Theta},t)}{\partial C(0)} &= \frac{K^2\exp\left(rt\right)}{(C(0)(\exp\left(rt\right)-1)+K)^2},\nonumber\\
\frac{\partial \mu(\boldsymbol{\Theta},t)}{\partial K} &= \frac{C(0)^2\exp\left(rt\right)(\exp\left(rt\right)-1)}{(C(0)(\exp\left(rt\right)-1)+K)^2}.\label{eq:LogisticPartials}
\end{align}

Recall that $\boldsymbol{\theta}$ includes only the unknown parameters to be estimated, so the components required from Equation (\ref{eq:LogisticPartials}) to form $\mathbf{J}(\boldsymbol{\theta})$ depend on the specific example.

Example synthetic logistic data is presented in Figure \ref{fig:LogisticFit}, demonstrating the model fits for $\boldsymbol{\theta} = (r,C(0))$, $\boldsymbol{\theta} = (r,K)$ and $\boldsymbol{\theta} = (r,\sigma)$. With data at \textit{early}, \textit{mid} and \textit{late} time, $T = (t_1,t_2,t_3) = (2.74,6.84,10.95)$ years, we observe an excellent model fit in all cases. The fit is best when $\boldsymbol{\theta} = (r,\sigma)$, as only one model parameter is unknown. Comparing $\boldsymbol{\theta} = (r,C(0))$ and $\boldsymbol{\theta} = (r,K)$ we observe a marginally better fit at late time when $K$ is known, and at early time when $C$ is known, as expected. 

\begin{figure}
	\centering
	\includegraphics[width=\linewidth]{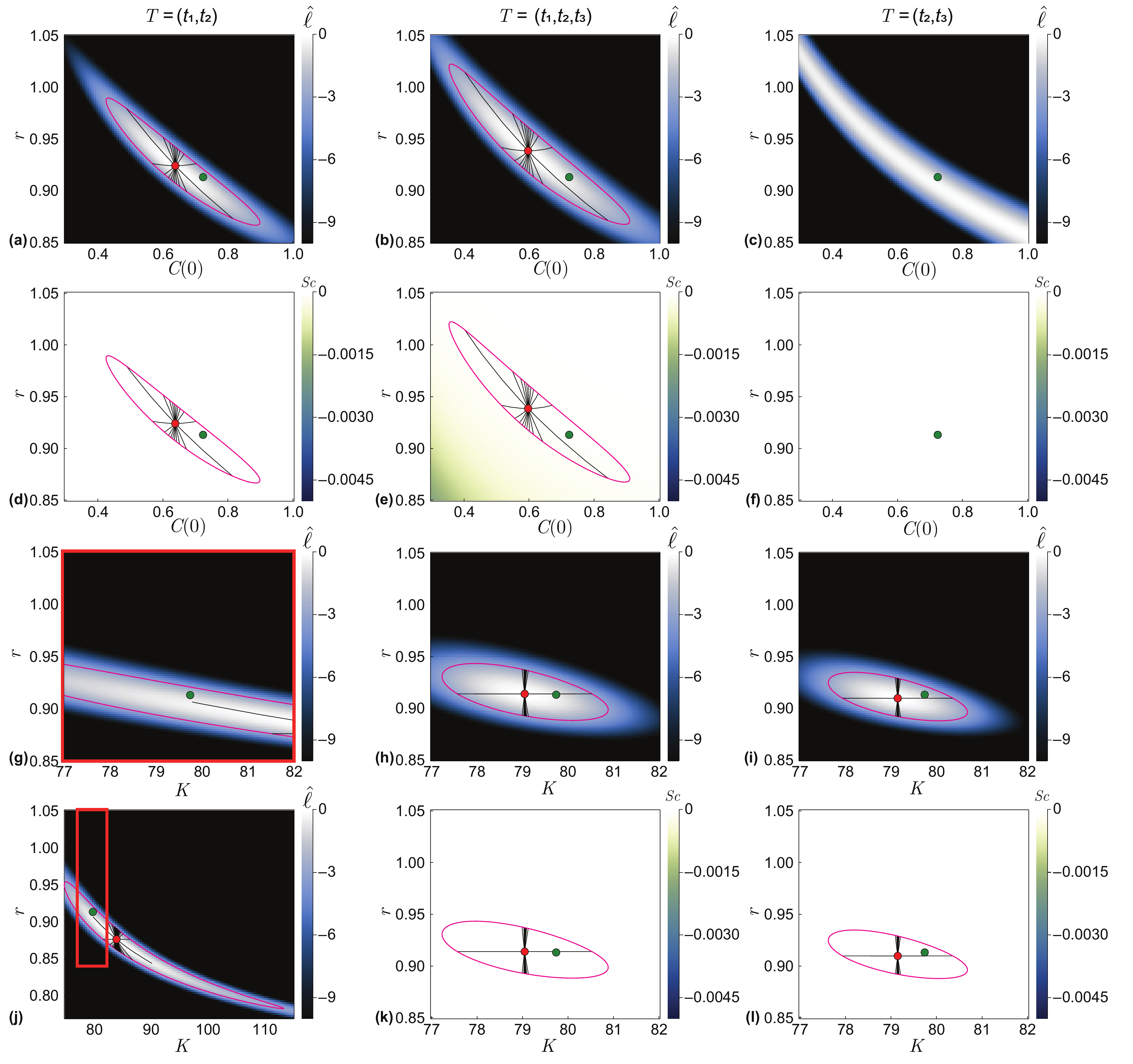}
	\caption{Logistic growth model with inferred growth rate, $r$, and initial condition, $C(0)$, (a-f); and with inferred growth rate, $r$, and carrying capacity, $K$, (g-l). True parameters are as noted in Figure \ref{fig:LogisticFit}, with known standard deviation, $\sigma = 2.301$. Heatmaps visualise the normalised log-likelihood (a-c,g-j) and the scalar curvature (d-f,k-l). The true parameter values are marked with green discs, with the MLEs indicated using red discs. Magenta curves correspond to likelihood-based 95\% confidence regions. Black lines are geodesic curves emanating from the MLEs, with lengths corresponding to a theoretical 95\% confidence distance. Columns of the figure correspond to observations from early-mid time ($T = (t_1,t_2)$), early-mid-late time ($T = (t_1,t_2,t_3)$), and mid-late time ($T = (t_2,t_3)$), where $(t_1,t_2,t_3) = (2.74, 6.84, 10.95)$ years. Each plot reflects a total of 30 observations, distributed equally between the specified time-points. The red outline in (j) corresponds to the (zoomed in) region (g), also outlined in red. In (g,j) we plot 1000 geodesics to observe the geodesic near the true parameter values.   We do not present $Sc$ corresponding to (g,j), however it is zero everywhere. \label{fig:Logistic12panel}}
\end{figure}

We present inference results for the logistic model for $\boldsymbol{\theta} = (r,C(0))$ in Figure \ref{fig:Logistic12panel}(a-f) and for $\boldsymbol{\theta} = (r,K)$ in Figure \ref{fig:Logistic12panel}(g-l). We do not present further results of inferring $\sigma$ for the logistic model, as little insight is gained beyond what we glean from the linear and exponential growth results. For $\boldsymbol{\theta} = (r,C(0))$, the normalised log-likelihood reflects the same relationship between growth rate and initial condition as for the linear and exponential case. With early-mid time data and early-mid-late time data, we are able to infer $\boldsymbol{\theta} = (r,C(0))$. With only mid-late time data, we find that the parameters are not practically identifiable. This can be seen from Figure \ref{fig:Logistic12panel}(c); the normalised log-likelihood remains above the threshold prescribed in Equation (\ref{eq:LikelihoodConfRegion}), and a closed likelihood-based 95\% confidence region cannot be constructed. \cbl This is also reflected in Figure \ref{fig:Logistic12panel}(f) alongside zero scalar curvature; such that the plot appears empty. \cb Comparing Figure \ref{fig:Logistic12panel}(a,b), and noting that they each rely on the same total number of observations, the importance of early and mid time data when inferring $\boldsymbol{\theta} = (r,C(0))$ is reinforced. The confidence region is tighter with only early-mid data, than with the same amount of data spread across early, mid and late times.      

Inferring $\boldsymbol{\theta} = (r,K)$ reflects similar behaviour. In Figure \ref{fig:Logistic12panel}(j) and associated zoomed-in view (Figure \ref{fig:Logistic12panel}g), inferring the carrying capacity from only early-mid time data results in an extremely wide confidence region, though the parameters remain identifiable. The geodesics emanating from the MLE match the likelihood-based confidence region very well in directions where the normalised log-likelihood is steep, however they do not quite reach the true parameter value in the direction where the normalised log-likelihood is relatively flat. Comparing Figure \ref{fig:Logistic12panel}(g,j) to Figure \ref{fig:Logistic12panel}(h,i), the MLE for $\boldsymbol{\theta} = (r,K)$ appears to be relatively poor when only early-mid time data is used.

When considering $\boldsymbol{\theta} = (r,C(0))$, we see that with early-mid time data and mid-late time data, the scalar curvature is zero everywhere. However, introducing a third time-point (early-mid-late data) results in a non-constant negative scalar curvature. We expect that this relates to the relationships between the parameters, and the difference between a mapping (where we have two pieces of information and two parameters to estimate), and a fit (where we have three pieces of information and two parameters to estimate). We do not observe similar behaviour for $\boldsymbol{\theta}=(r,K)$ with data at three time-points; the scalar curvature still appears to be zero everywhere.  One explanation for this is that data at $t_1$, where $C(t) \ll K$, may be effectively independent of $K$; providing no information about $K$ \cite{Warne2017}. This may effectively reduce the problem to a mapping. Given that the scalar curvature is a feature of the manifold rather than the data, it is of interest to investigate what would happen, were the true parameters to lie within this region of non-constant scalar curvature.

\begin{figure}[h!]
	\centering
	\includegraphics[width=0.5\linewidth]{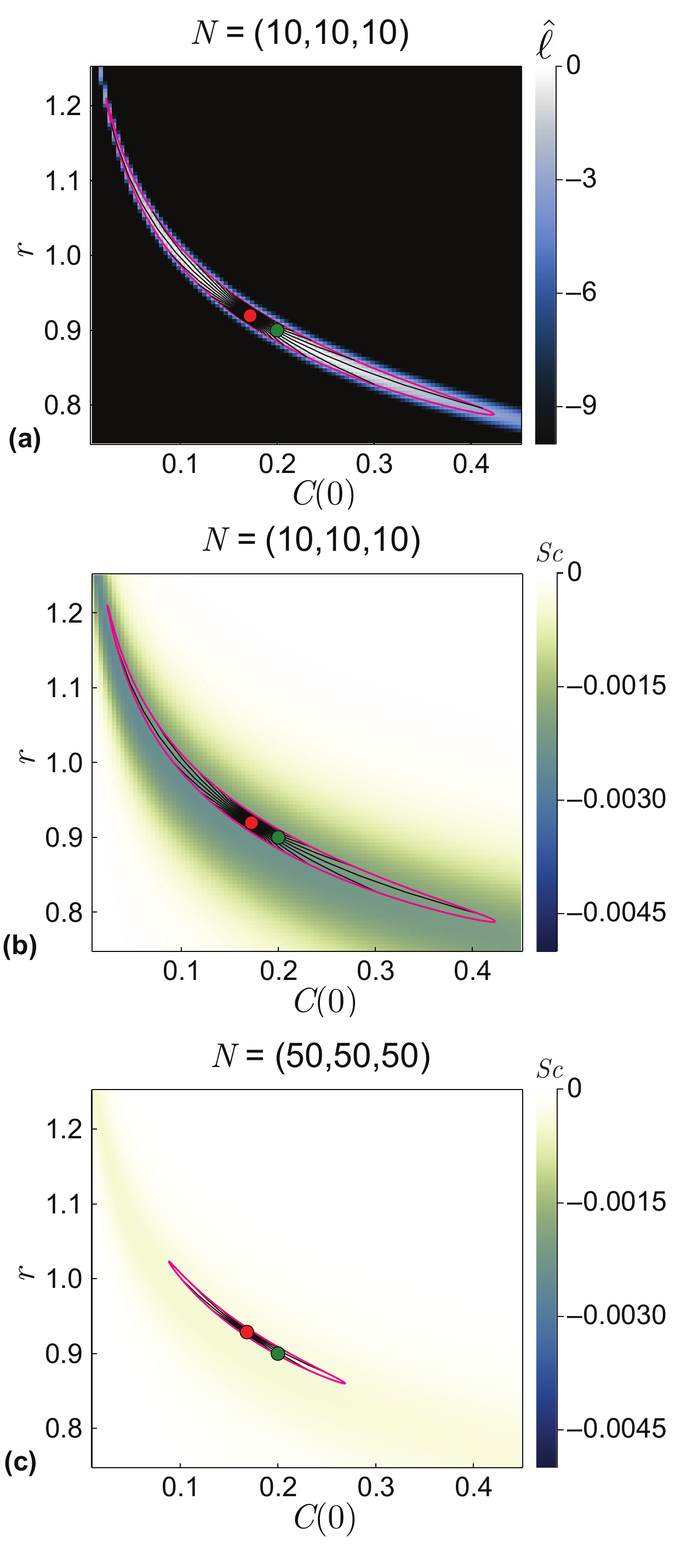}
	\caption{Logistic growth model with inferred growth rate, $r$, and initial condition, $C(0)$, with known standard deviation, $\sigma = 2.301$ and carrying capacity, $K = 79.74$. Heatmaps visualise the normalised log-likelihood (a) and the scalar curvature (b-c). Data is observed at $T = (2.74,6.84,10.95)$, with 10 (a-b) and 50 (c) observations per time-point.   The true parameter values are marked with green discs, with MLEs indicated using red discs. Magenta curves correspond to likelihood-based 95\% confidence regions. Black lines are 100 geodesic curves emanating from the MLEs, with lengths corresponding to a theoretical 95\% confidence distance. \label{fig:HighCurvature}}
\end{figure}

To address this, we generate an alternate set of synthetic logistic growth data using parameter values from within the high curvature region;\\ $(r,C(0)) = (0.9,0.2)$, with $(K,\sigma) = (79.74,2.301)$ as before. Inference results are presented in Figure \ref{fig:HighCurvature}. We still observe correspondence between the endpoints of the geodesics and the likelihood-based confidence region, however the confidence region is now significantly narrower and reflects a more hyperbolic shaped relationship between $r$ and $C(0)$ in terms of the normalised log-likelihood. Increasing the number of observations, as depicted in Figure \ref{fig:HighCurvature}c, has the expected effects of tightening the confidence region and reducing the scalar curvature. This reduces the apparent curvature of the confidence region.

\FloatBarrier

\subsection{SIR epidemic model}

The SIR model describes the dynamics of epidemic transmission through a population \cite{Murray2002}. Populations are assumed to be comprised of susceptible, $s(t)$, infected, $i(t)$, and recovered, $r(t)$, individuals. The total population, $\mathscr{N}$, is held constant. When analysing the SIR model in this work, we consider each population as a proportion of the total population, such that $S(t) = s(t)/\mathscr{N}$, $I(t) = i(t)/\mathscr{N}$, and $R(t) = r(t)/\mathscr{N}$.  Quantities $\mathscr{N}$, $s(t)$, $i(t)$ and $r(t)$ are dimensional with dimensions of number of individuals, whereas $S(t) \in [0,1]$, $I(t) \in [0,1]$ and $R(t) \in [0,1]$ are dimensionless quantities with the property that $S(t)+I(t)+R(t)=1$. While the coral re-growth process considered in the population model examples takes place over many years, epidemic occur over a timescale of days or weeks. As such, we now take $t$ to represent time as measured in days, rather than years. The parameters of the SIR model are the infection rate, $\beta$ [day$^{-1}$], and the rate at which infected individuals are removed, $\gamma$ [day$^{-1}$], for example, via recovery from the infection: 
\begin{align}
\frac{\textrm{d}S}{\textrm{d}t} &= -\beta S I,\nonumber \\
\frac{\textrm{d}I}{\textrm{d}t} &= \beta S I -\gamma I,\nonumber \\
\frac{\textrm{d}R}{\textrm{d}t} &= \gamma I.\label{eq:SIRmodel}
\end{align} 

Alongside $\beta$ and $\gamma$ we could also treat the initial conditions; $S(0)$, $I(0)$ and $R(0)$, as unknown parameters to be estimated.  The standard SIR model presented in Equation (\ref{eq:SIRmodel}) is sufficient for our purposes in this work, however numerous extensions to the SIR model are considered in the literature. These extensions incorporate factors such as age structure, birth and death, exposed but not yet infected individuals, seasonality, competition between infectious strains,  waning immunity, vaccination and spatial structure \cite{Carcione2020,Murray2002,Papst2019,Roberts2007}.  
\begin{figure}
	\centering
	\includegraphics[width=\linewidth]{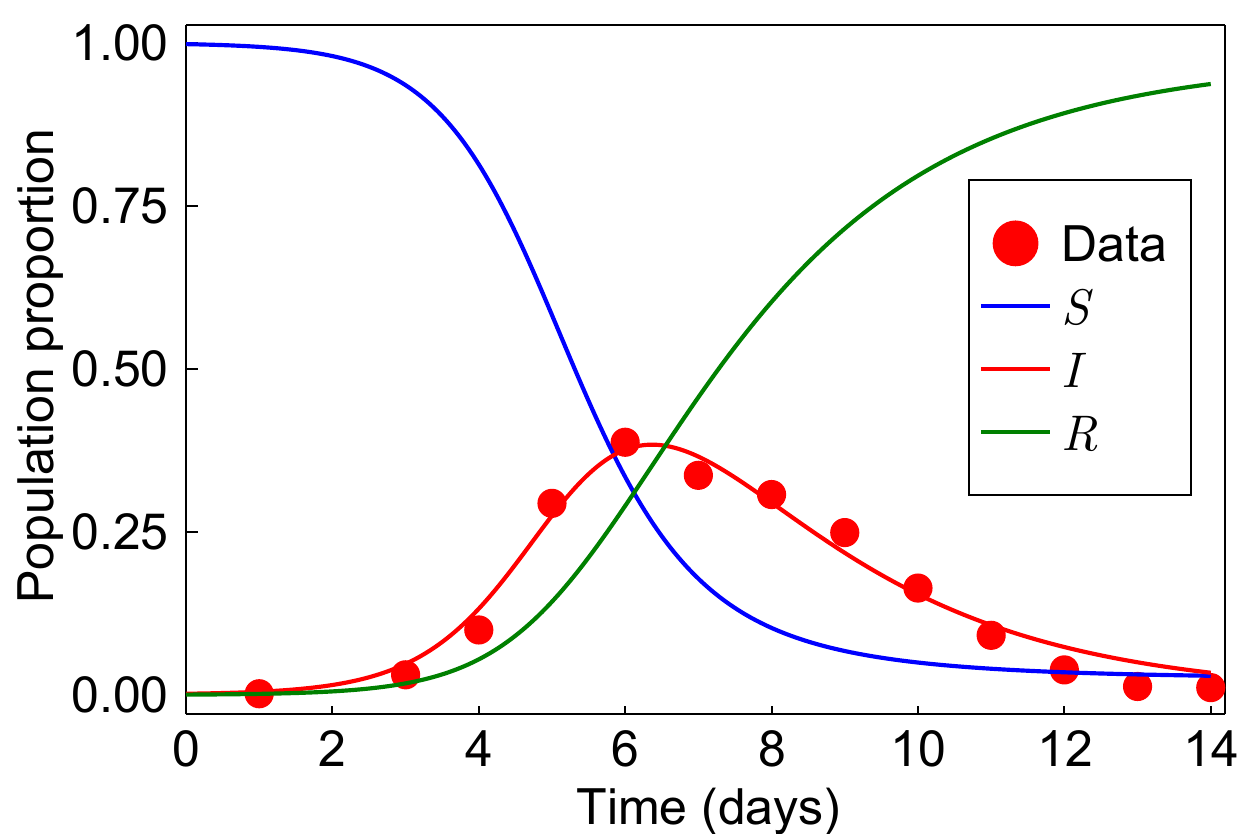}
	\caption{Data marked with red discs represents number of infected individuals during an influenza outbreak in a boarding school \cite{Murray2002}. Susceptible, $S(t)$, infected, $I(t)$, and recovered, $R(t)$ populations are modelled according to Equation (\ref{eq:SIRmodel}) based on parameters inferred in \cite{Murray2002}, $\beta = 1.6633$, $\gamma = 0.44036$; which we treat as the true parameters when generating synthetic data. Initial population proportions are $S(0) = 762/763$, $I(0) = 1/763$ and $R(0) = 0$. \label{fig:SIRData}}
\end{figure}

\begin{figure}[h]
	\centering
	\includegraphics[width=\linewidth]{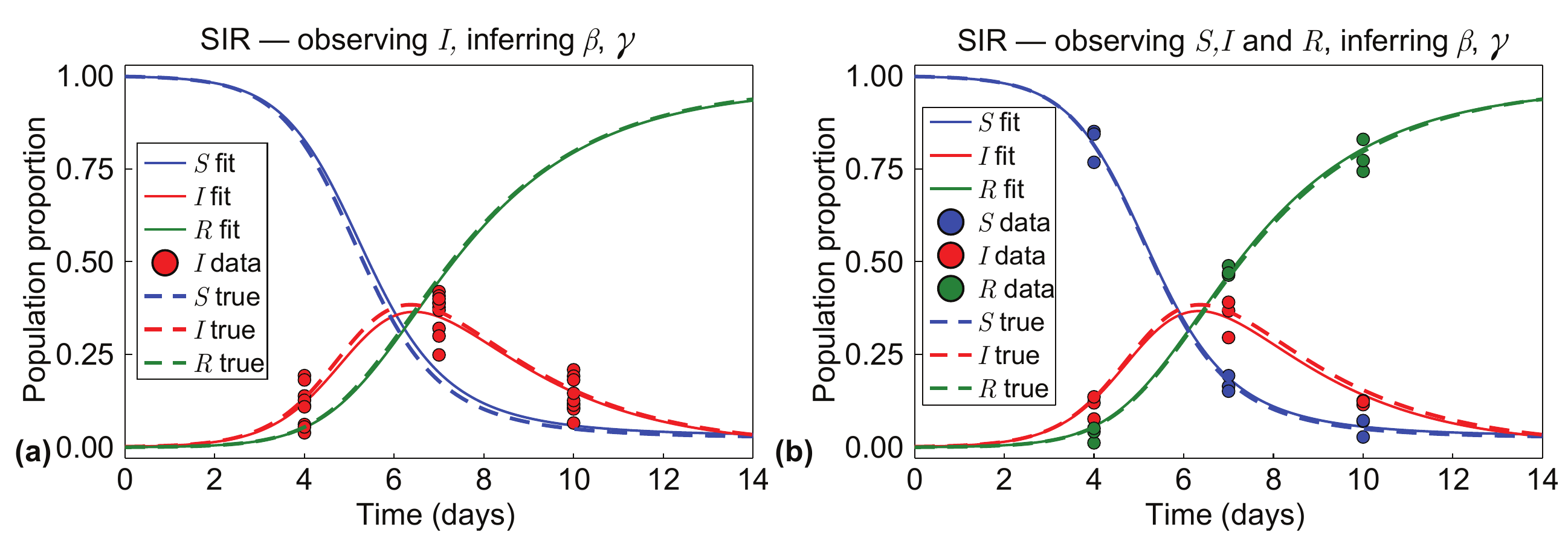}
	\caption{Example synthetic data generated from the SIR model under the scenarios where: (a) only the number of infected individuals is observed, and; (b) where we have observations pertaining to all three populations. Observations are marked with discs. Populations are modelled according to Equation (\ref{eq:SIRmodel}) based on parameters inferred in \cite{Murray2002}. Initial population proportions are $S(0) = 762/763$, $I(0) = 1/763$ and $R(0) = 0$. In (a) there are $N = 10$ observations at each time-point, and we prescribe $\sigma = 0.05$; in (b) there are three observations per time-point, per population, with prescribed $\sigma = 0.03$. The choices of $\sigma$ are sufficiently small that the data generated consists only of positive observed population proportions. \label{fig:SIRSyntheticData}}
\end{figure}

Data pertaining to the proportion of a population infected during an influenza outbreak in a boarding school is presented in Figure \ref{fig:SIRData}. Observations in the original data record the number of infected individuals over a 14 day period \cite{Murray2002}, in a population of $\mathscr{N} = 763$, with initial populations $(s(0),i(0),r(0)) = (762,1,0)$. This data is used in \cite{Murray2002} to estimate parameters for the SIR model, which, after scaling such that $S+I+R= 1$, are $\beta = 1.6633$ and $\gamma = 0.44036$. We treat these values as the true parameters when generating synthetic data; examples of which are presented in Figure \ref{fig:SIRSyntheticData}. In the context of an SIR model, the presence of multiple observations at a single time-point could reflect, for example; reporting errors, uncertainty in test accuracy or expert judgement \cite{Hay2021,Zimmer2018}. In the boarding school data considered in \cite{Murray2002}, observations pertain only to the number of infected individuals. Given that the SIR model features multiple populations, data could in theory contain observations of the other populations also. Example synthetic data with observations on all three populations is presented in Figure \ref{fig:SIRSyntheticData}b.

The SIR model as described in Equation (\ref{eq:SIRmodel}) does not admit a closed form analytical solution, so we apply numerical techniques to solve the system. This becomes somewhat computationally expensive, as the Fisher information computations rely on partial derivatives of the model solution with respect to the parameters to form the model Jacobian, and the information geometry computations require partial derivatives of the Fisher information up to second order. Approximating these partial derivatives using numerical techniques entails solving the system of ODEs several times. Some computational cost may be spared through taking advantage of the known relationship that $S+I+R = 1$.  

\begin{figure}
	\centering
	\includegraphics[width=\linewidth]{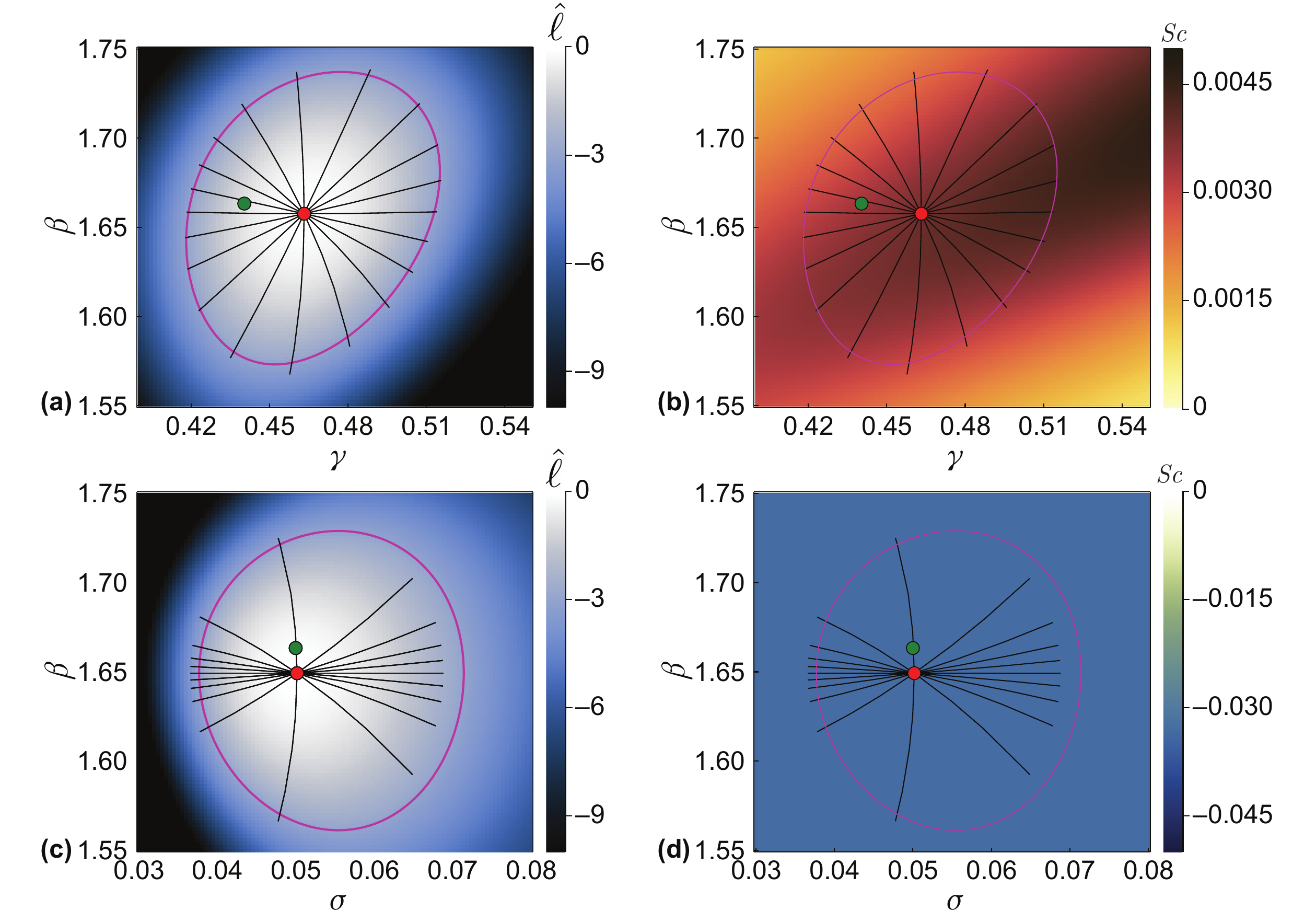}
	\caption{Inferring $\boldsymbol{\theta} = (\beta,\gamma$) in (a,b) and $\boldsymbol{\theta} = (\beta,\sigma)$ in (c,d), for the SIR model with observations only on the number of infected individuals. Observations in the synthetic data occur at $T = (4,7,10)$, with $N = 10$ observations per time-point. True parameters, $(\beta,\gamma,\sigma) = (1.66334,0.44036,0.05)$, are marked with green discs, with MLEs indicated using red discs. Magenta curves correspond to likelihood-based 95\% confidence regions. Black lines are geodesic curves emanating from the MLEs, with lengths corresponding to a theoretical 95\% confidence distance.  Initial populations are as described in Figure \ref{fig:SIRSyntheticData}.  \label{fig:SIRObsIonly} }
\end{figure}

\begin{figure}
	\centering
	\includegraphics[width=\linewidth]{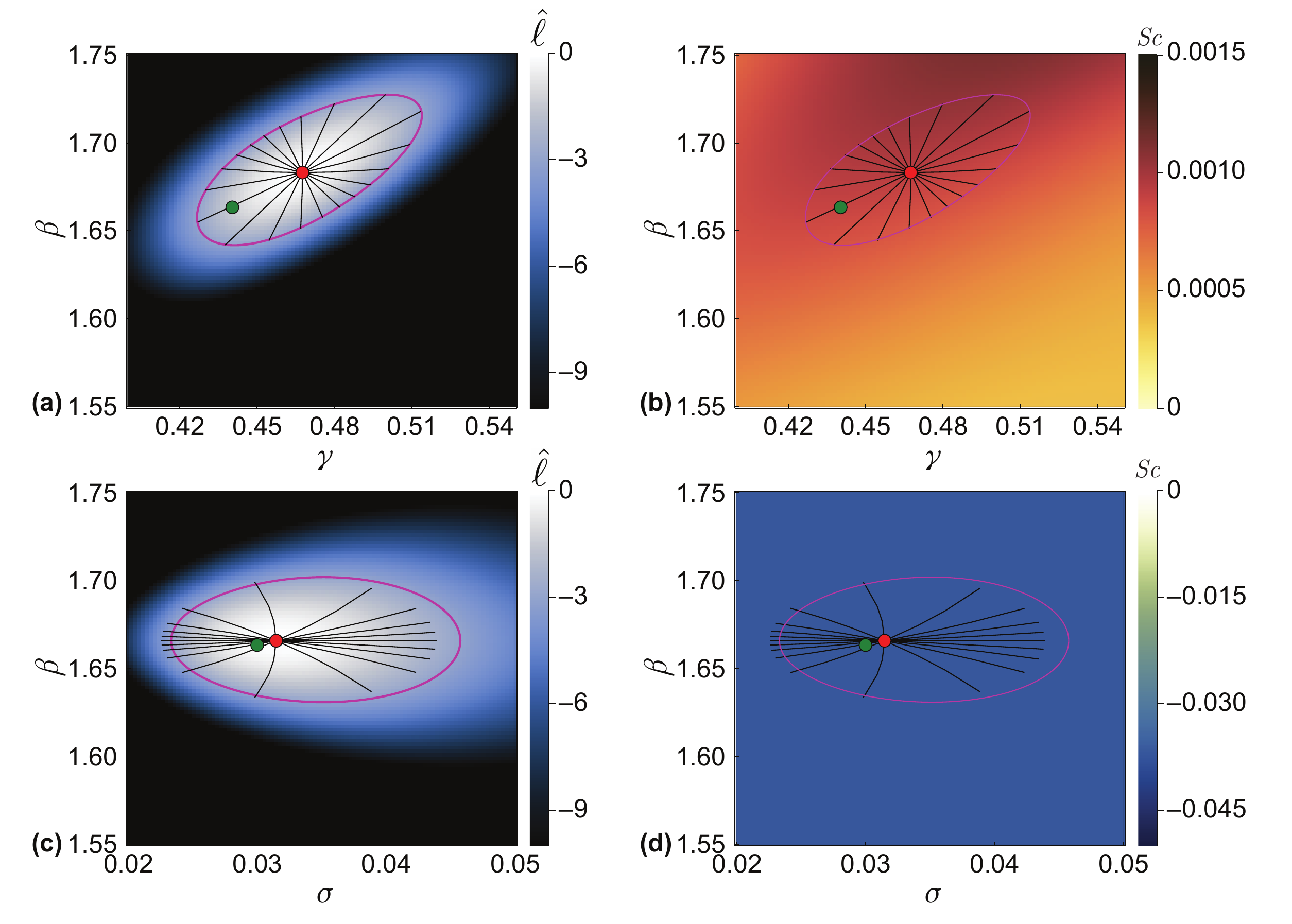}
	\caption{Inferring $\boldsymbol{\theta} = (\beta,\gamma$) in (a,b) and $\boldsymbol{\theta} = (\beta,\sigma)$ in (c,d), with observations on all three variables, $S$, $I$ and $R$. Observations in the synthetic data occur at $T = (4,7,10)$, with three observations of each population at each time-point; 27 observations in total, as depicted in Figure \ref{fig:SIRSyntheticData}b. True parameters,  $(\beta,\gamma,\sigma) = (1.66334,0.44036,0.03)$, are marked with green discs, with MLEs indicated using red discs. Magenta curves correspond to likelihood-based 95\% confidence regions. Black lines are geodesic curves emanating from the MLEs, with lengths corresponding to a theoretical 95\% confidence distance. Initial populations as described in Figure \ref{fig:SIRSyntheticData}. \label{fig:SIRObsAll}  }
\end{figure}

For brevity, we restrict our investigation of the SIR model to the cases where $\boldsymbol{\theta} = (\beta,\gamma)$ and $\boldsymbol{\theta} = (\beta,\sigma)$. Results in Figure \ref{fig:SIRObsIonly} correspond to the case where observations pertain only to the number of infected individuals, while those in Figure \ref{fig:SIRObsAll} are produced from data containing observations of all three populations. In both cases the results for $\boldsymbol{\theta} = (\beta,\sigma)$ align with that observed in previous results; the geodesics appear to define a marginally smaller area and are offset from the likelihood-based confidence regions in the direction of decreasing $\sigma$, and the scalar curvature is the constant $Sc = -1/N$.       

Regardless of whether we observe only the infected population or all populations, inferring $\boldsymbol{\theta} = (\beta,\gamma)$ produces a non-constant positive scalar curvature. In Figure \ref{fig:SIRObsIonly}b, where only $I$ is observed, we see that the geodesics emanating from the MLE extend beyond the likelihood-based confidence region. This also occurs in Figure \ref{fig:SIRObsAll}b, where all three populations are observed, however it is difficult to perceive at this scale. Based on this result, and the observations involving negative scalar curvature when inferring $\sigma$, it might seem that positive scalar curvature produces geodesics that extend beyond corresponding likelihood-based confidence regions, whereas negative scalar curvature has the opposite effect. However, repeating the analysis with different synthetic data sets---generated from a different random seed---suggests that in some cases the geodesics will extend beyond the likelihood-based confidence regions, and in some cases they will fall short, however the scalar curvature remains positive in all cases.

\clearpage
\subsection{Hypothesis testing}

\begin{figure}[h!]
	\centering
	\includegraphics[width=\linewidth]{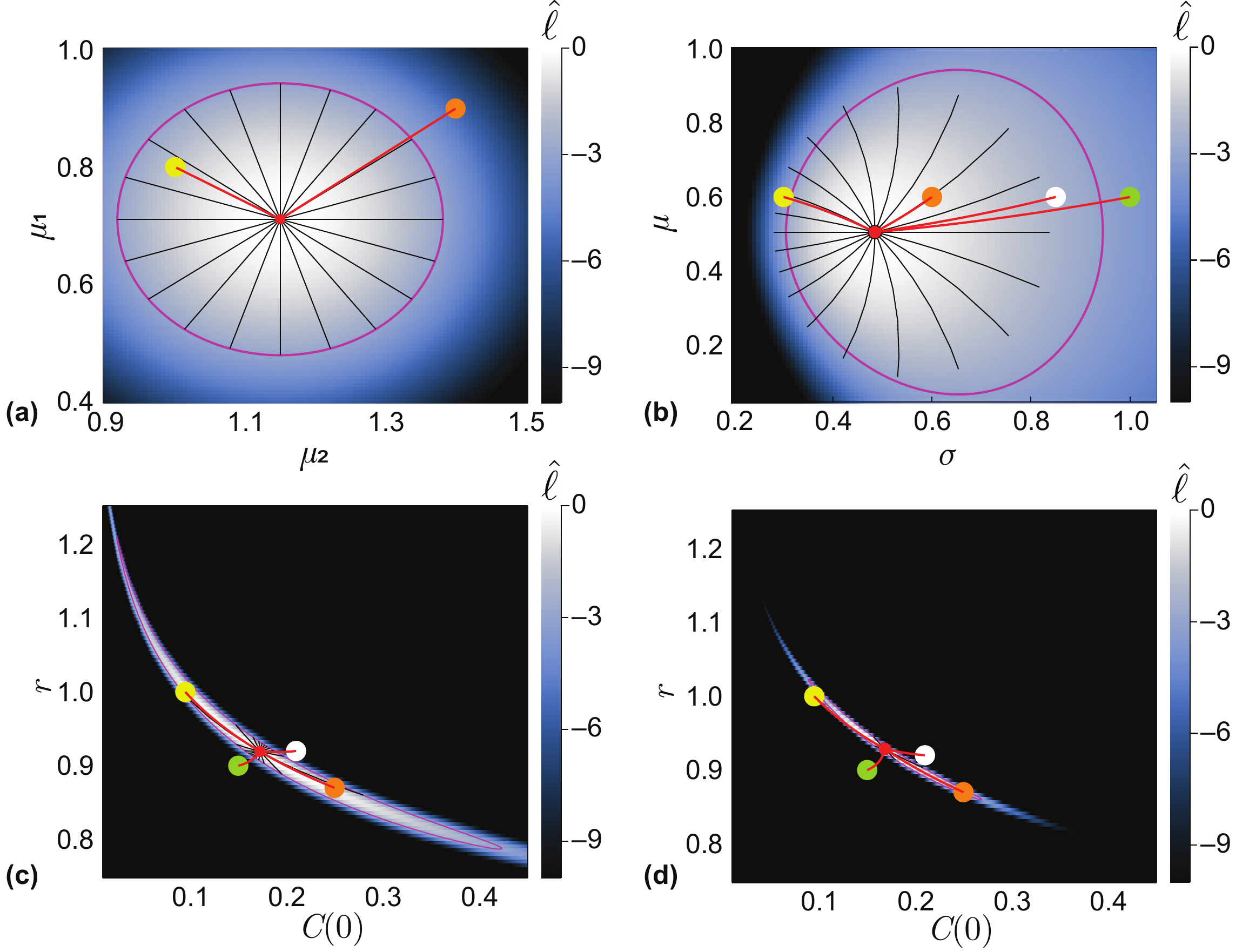}
	\caption{Example hypothesis tests for the: (a) univariate normal distribution, with $\boldsymbol{\theta} = (\mu,\sigma)$, $\hat{\boldsymbol{\theta}}=  (0.5050,0.4846)$; (b) multivariate normal distribution, with $\boldsymbol{\theta} = (\mu_1,\mu_2)$, $\hat{\boldsymbol{\theta}}=  (0.7109,1.1498)$; logistic model with $\boldsymbol{\theta} = (r,C(0))$ in the high curvature region as considered in Figure \ref{fig:HighCurvature}, with (c) $N = (10,10,10)$, $\hat{\boldsymbol{\theta}}= (0.9195,0.1723)$, and (d) $N = (50,50,50)$, $\hat{\boldsymbol{\theta}}= (0.9287,0.1682)$. In each case, we test several example hypotheses, $\boldsymbol{\theta}_0$, marked by coloured discs. Geodesics between the MLEs (red discs) and each $\boldsymbol{\theta}_0$ are shown in red.  Magenta curves correspond to likelihood-based 95\% confidence regions. Black lines are geodesic curves emanating from the MLEs, with lengths corresponding to a theoretical 95\% confidence distance.  \label{fig:HypothesisTests}}
\end{figure}

In Figure \ref{fig:HypothesisTests} we present several example hypothesis tests, using both likelihood-ratio-based and geodesic-distance-based approaches, as outlined in Section \ref{sec:Methods}. Test statistics and corresponding $p$-values for each hypothesis test are provided in Table 1. For the multivariate normal distribution, where we observe that the endpoints of geodesics corresponding to a theoretical 95\% confidence distance align closely with the likelihood-based 95\% confidence regions, we find that the results of the hypothesis tests are near-identical. Further, the hypothesis test results are consistent with our interpretation of the 95\% confidence regions; test points within the confidence regions have $p$-values greater than 0.05, while test points outside the confidence regions have $p$-values less than 0.05. 

We also perform hypothesis tests for the logistic model in the high curvature region of parameter space. Like before, results are comparable for different numbers of observations at each time point; $N = (10,10,10)$ and $N = (50,50,50)$, as considered in Figure \ref{fig:HighCurvature}. Even in this high curvature region, we find that the endpoints of geodesics corresponding to a theoretical 95\% confidence distance very closely match the likelihood-based 95\% confidence regions. This is again reflected in the results of the hypothesis tests, where very similar results are obtained from the likelihood-ratio-based hypothesis tests and the geodesic-distance-based hypothesis tests, even for relatively extreme $\boldsymbol{\theta}_0$. As the number of observations increases, we observe for each  $\boldsymbol{\theta}_0$ considered, that in accordance with the confidence regions tightening, the test statistics increase and accordingly $p$-values decrease. 
 
\begin{table}[h] 
	\centering \textbf{\captionof{table}{Hypothesis test results}} 
\begin{tabular}{|l|c|c|c|c|c|} 
\hline Model  & $\boldsymbol{\theta}_0$ & $ \lambda_{LR} $ & $\lambda_{GD}$ & $p_{LR}$ & $p_{GD}$\\ 
\hline 
Multivariate&(0.8,1.0)& 3.3737&3.3737&0.1851&0.1851\\
Normal&(0.9,1.4)&10.9297& 10.9297&0.0042&0.0042\\\hline
Univariate&(0.6,0.3)& 7.5051&5.1954& 0.0235&0.0744\\
Normal&(0.6,0.6)& 1.0460&1.2201&0.5927&0.5433\\
&(0.6,0.85)&4.6134&6.5226&0.0996&0.0383\\
&(0.6,1.0)&6.9271&10.6665&0.0313&0.0048\\\hline
Logistic&(1.0,0.095)&2.7086&2.5798&0.2581&0.2753\\
$(10,10,10)$&(0.87,0.25)&1.6387&1.5626& 0.4407&0.4578\\
&(0.92,0.21)&30.0130&29.9821&3.0391$\times10^{-7}$&3.0865$\times10^{-7}$\\
&(0.9,0.15)&56.2776&56.5328&6.0185$\times10^{-13}$&5.2969$\times10^{-13}$\\\hline
Logistic&(1.0,0.095)&31.3038&31.0222&1.5939$\times10^{-7}$&1.8349$\times10^{-7}$\\
$(50,50,50)$&(0.87,0.25)&4.2062&4.2276&0.1221&0.1208\\
&(0.92,0.21)&97.6247&97.5164&$<10^{-16}$&$<10^{-16}$\\
&(0.9,0.15)&368.1479&368.7335&$<10^{-16}$&$<10^{-16}$\\
\hline
\end{tabular}
\end{table}

\begin{figure}[h]
	\centering
	\includegraphics[width=\linewidth]{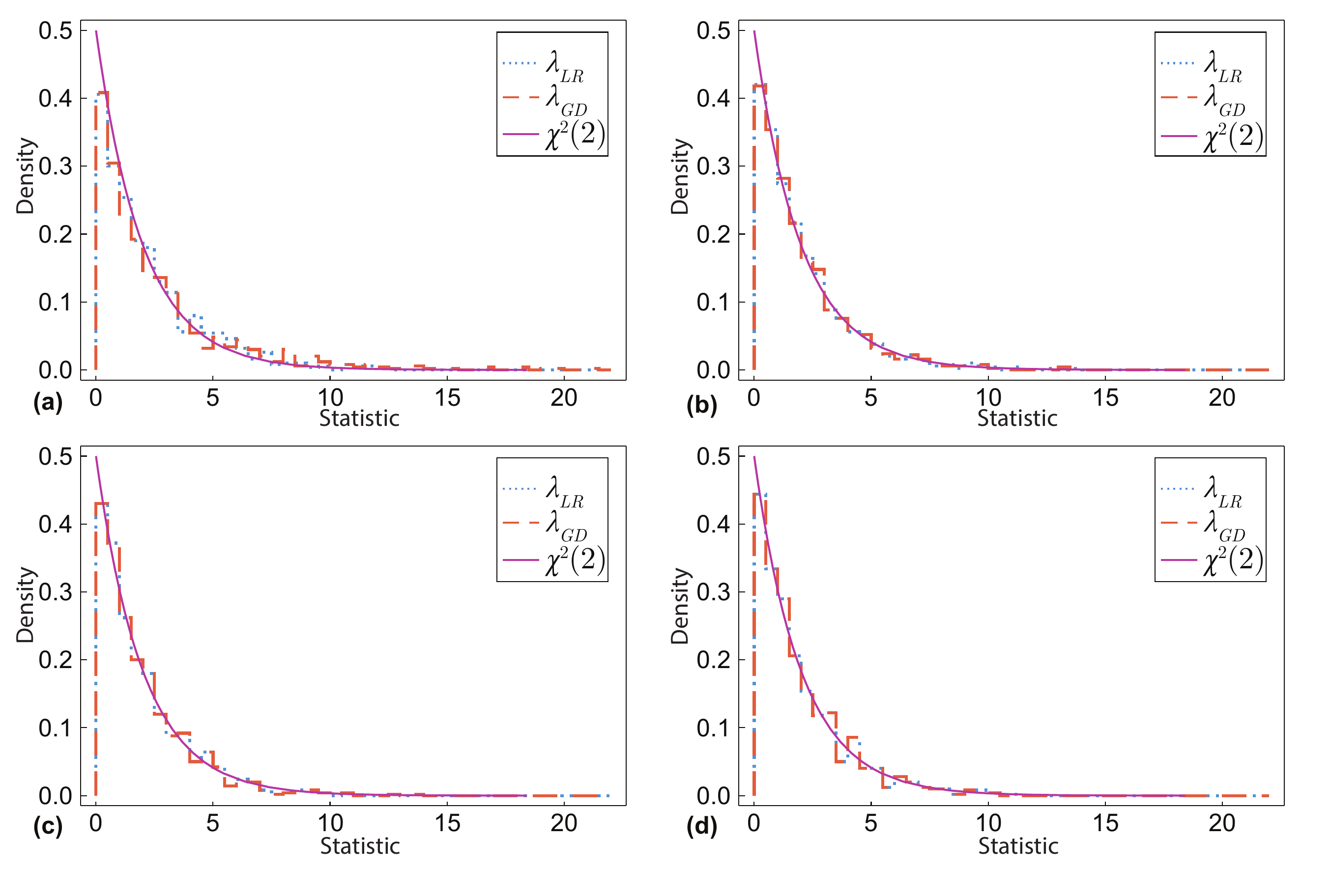}
	\caption{Step histograms show the density of the distribution of test statistics for each hypothesis testing approach, for (a,b): the univariate normal distribution with $\boldsymbol{\theta} = (\mu,\sigma)$, and (c,d): the multivariate normal distributions with $\boldsymbol{\theta} = (\mu_1,\mu_2)$. Test statistics are computed from the true parameter values and the MLE, for 1000 sets of synthetic data. Datasets represented in (a,c) contain $N=10$ observations, while in (b,d) $N = 1000$. Purple curves correspond to the density of the $\chi^2(2)$ distribution, while blue dotted lines represent the likelihood-ratio-based test statistics and orange dashed lines represent the geodesic-distance-based test statistics. \label{fig:HT_Curves}}
\end{figure}

As we are using synthetic data and know the true parameters, we can use hypothesis testing to pedagogically investigate Wilks' theorem \cite{Pawitan2001} and the asymptotic relationship given in (\ref{eq:geoCRdist}). We generate 1000 synthetic datasets and for each dataset perform a hypothesis test for the true parameters. This is repeated for the univariate and multivariate normal distributions with $N=10$ and $N=1000$ observations. In Figure \ref{fig:HT_Curves} we present densities for both the likelihood-ratio-based and geodesic-distance-based test statistics, alongside the probability density of $\chi^2(2)$. For the multivariate normal distribution with $\boldsymbol{\theta} = (\mu_1,\mu_2)$, the density profiles for $\lambda_{LR}$ and $\lambda_{GD}$ are near-identical; as expected following the results in Figure \ref{fig:HypothesisTests} and Table 1. We also observe a good match between these profiles and $\chi^2(2)$, even with just $N = 10$. For the univariate normal distribution with $\boldsymbol{\theta} = (\mu,\sigma)$, when $N = 10$ we observe differences between  $\lambda_{LR}$ and $\lambda_{GD}$. Both profiles are similar to $\chi^2(2)$, though there appears to be a higher density in the tails of the distributions of the test statistics. As the number of observations increases to $N=1000$, the difference between $\lambda_{LR}$ and $\lambda_{GD}$ reduces significantly, and both closely match $\chi^2(2)$. 

From Wilks' theorem \cite{Pawitan2001} and (\ref{eq:geoCRdist}), asymptotically 95\% of the 95\% confidence regions we construct should contain the true parameter values. We can determine what proportion of the likelihood-based and geodesic-distance-based 95\% confidence regions that we construct contain the true parameter values using the information presented in Figure \ref{fig:HT_Curves}. This is done by comparing the test statistics to the critical value; $\Delta_{2,0.05}$, from (\ref{eq:LikelihoodConfRegion}). For the multivariate normal distribution with $N = 10$ we find that 95.7\% of the likelihood-based and geodesic-distance-based confidence regions contain the true parameter values. With $N = 1000$ we find that 94.8\% contain the true parameters, approaching the theoretical 95\%. For the univariate normal distribution with $N = 10$ we find that 93.2\% of the likelihood-based confidence regions contain the true parameter, while  only 88.0\% of the geodesic-distance-based confidence regions contain the true parameters. With $N = 1000$, we find that 95.2\% of the likelihood-based confidence regions and 95.1\% of the geodesic confidence regions contain the true parameters.            

\FloatBarrier
\clearpage
\section{Discussion}\label{sec:Discussion}

Parameter estimation is wrought with challenges relating to the availability and quality of experimental or field data \cite{Frolich2014,Hlavacek2011,Toni2009,Vo2015}. This prompts a strong consideration of uncertainty quantification to support point-estimation of model parameters \cite{DevenishNelson2010}. In this section, we discuss the results presented in Section \ref{sec:Results}. We highlight opportunities for application of information geometry techniques, including geodesic curves and scalar curvature; to supplement traditional maximum likelihood based parameter inference and uncertainty quantification. We conclude by outlining areas for further investigation.

Even for relatively small sample sizes, we observe good correspondence between the likelihood-based 95\% confidence regions and the end-points of geodesic curves corresponding to a theoretical 95\% confidence distance, in accordance with the asymptotic relationship described in Equation (\ref{eq:geoCRdist}); particularly when estimating model parameters. When estimating standard deviation, as outlined in Section \ref{sec:Results}, geodesics appear to suggest a tighter confidence region, and appear to be biased towards parameter space with smaller standard deviation. We observe this effect decreasing as the number of observations increases; in line with the known underestimation bias of minimum likelihood estimates of variance \cite{Pawitan2001}. The misalignment of likelihood-based confidence regions and geodesic endpoints appears to occur more frequently in examples with non-zero scalar curvature, although we observe a good match in Figure \ref{fig:HighCurvature} despite the non-constant scalar curvature. 

Visualising the scalar curvature throughout a parameter space can indicate areas where there may be issues with identifiability. Areas with significant non-constant scalar curvature can suggest a complicated relationship between parameters in terms of the normalised log-likelihood, such as the hyperbolic confidence region observed in Figure \ref{fig:HighCurvature}. However, it is possible to produce examples, such as Figure \ref{fig:Logistic12panel}(c,f), where there is practical non-identifiability despite zero scalar curvature everywhere. Although we do not show it here, for the logistic model with $\boldsymbol{\theta} = (r,K)$ in the region of parameter space where $C(0)\approx K$, computation of the scalar curvature breaks down as the Fisher information matrix becomes singular. Here, it may be obvious that we can not identify the growth rate, $r$, from a process that is initialised at its steady state ($C(0) = K$). However, observing this behaviour in general may help to detect issues with identifiability, particularly for models without analytical solutions.   

The information geometry techniques we discuss are primarily implemented numerically; as such there is a computational cost to consider. For the normal distributions and population growth models in this work, where analytical solutions are available, the information geometry techniques are not disproportionately more computationally expensive than the traditional likelihood-based inference and confidence regions. Examples such as the SIR model, where no analytical solution is available, represent a significantly greater computational burden. However, this impacts both the likelihood based inference and information geometry techniques as the underlying system of ODEs, for example Equation (\ref{eq:SIRmodel}), must be solved numerous times. The computational cost associated with the information geometry techniques depends significantly on the desired resolution for the scalar curvature surface, and on the number of geodesic curves. A suitable approach may be to first compute the scalar curvature on a coarse grid to identify areas of interest to investigate with a refined grid. \cbl Further, the geodesic curves and scalar curvature computations are highly amenable to parallelisation, which can significantly reduce computation time. \cb

This computational cost will generally pale in comparison to the costs associated with collecting experimental or field data, \cbl and may be easily justified if the information geometry techniques are used to guide data collection. \cb If information geometric analysis identifies a region of parameter space with significant non-constant scalar curvature for a model, such as in Figure \ref{fig:HighCurvature}, and practitioners have a prior expectation that the true parameter values fall somewhere within this region, this may indicate that a greater quantity or quality of data is needed to improve identifiability for that particular model. Alternatively, such analysis may guide practitioners in choosing favourable experimental conditions; for example in cell culture experiments, where it is possible to vary the initial cell seeding density \cite{Browning2021a}. \cbl Experimental design is a process wherein experiments are performed or simulated iteratively with perturbations, such that some measure of information is maximised. Through this process the most informative experiments are identified, facilitating design of optimal experimental protocols \cite{Hagen2013,Liepe2013,Ruess2015}. Common to these approaches is the importance of quantifying and comparing information. While we do not consider optimal experimental design in this work, there is potential to incorporate information geometric techniques in the experimental design process as a means of comparing information between experimental perturbations. This is an area for further investigation.   \cb   

Although we focus on how information geometry can supplement traditional maximum likelihood based inference and uncertainty quantification, primarily through visualisation, it should be noted that concepts from information geometry have also found application in the inference context from a computational efficiency standpoint. For example in Bayesian inference, by defining Monte Carlo sampling methods on a Riemann manifold, the geometric structure of the parameter space can be exploited \cite{Girolami2011}. Simulated paths across the manifold automatically adapt to local structure, facilitating efficient convergence, even in higher dimensions and in the presence of strong correlation \cite{Girolami2011,House2016}. Concepts from information geometry, including geodesic curves, are also implemented in methods for model reduction \cite{Transtrum2014}. \cbl These applications of information geometry techniques to improve computational algorithms highlight further utility of geometric concepts for inference in higher dimensions, beyond that which we demonstrate through visualisation in this work. \cb  

Geodesics can be used to measure the distance between probability distributions. As demonstrated in Section \ref{sec:Results}, it is possible to perform hypothesis tests based on geodesic distance \cite{Kass1997,Menendez1995,Nielsen2020}. The approach for performing a hypothesis test is to solve a boundary value problem to find the geodesic connecting two points in parameter space, and use the corresponding geodesic distance to compute a test statistic. For the examples considered in this work, such boundary value problems are readily solved numerically using standard techniques, such as those included in the Julia package DifferentialEquations.jl \cite{Rackauckas2017}. Careful numerical handling may be required for geodesic curves close to boundaries of parameter space. For more complicated examples, particularly those in high-dimensional manifolds,  achieving converging solutions to geodesic boundary value problems can prove challenging. There is scope for a review of the different numerical methods for solving boundary value problems, with a particular focus on their applicability to solving geodesic boundary value problems for hypothesis testing in high-dimensional manifolds.   

\cbl In this work we only consider models that admit unimodal likelihoods. In cases where the likelihood is multimodal; provided that we are able to obtain the Fisher information required to compute the Christoffel symbols, we are still able to compute the scalar curvature and perform hypothesis tests based on geodesic distance. With multimodal likelihoods, it would not be possible to construct confidence regions from geodesics emanating from the MLE. Although, we note that constructing confidence regions for multimodal likelihoods is also problematic with traditional likelihood-based inference methods.  \cb

There are several avenues for future research in this area. Here, we consider two-dimensional manifolds to facilitate convenient visualisation, however the inference and information geometry techniques are general, and can be readily applied to higher dimensional manifolds \cite{Amari2000,Pawitan2001}; albeit with increased computational cost. Extending this analysis to three dimensions would enable consideration of situations where there is scalar curvature associated both with the variability of the observation process, $\sigma$, and also with interactions between model parameters; for example, it may be insightful to consider $\boldsymbol{\theta} =(\beta,\gamma,\sigma)$ for the SIR model, where we associate a constant negative scalar curvature with $\sigma$ and non-constant positive scalar curvature due to interactions between $\beta$ and $\gamma$. In three dimensions, likelihood-based confidence regions can be visualised as a series of two-dimensional slices oriented in three-dimensional space \cite{Browning2021a}; this technique could be applied to visualise slices of the scalar curvature in three-dimensions. One approach for visualisation in higher dimensions is to produce an ensemble of these two- or three-dimensional confidence regions for various combinations of parameters of interest, with other parameters fixed at their MLEs. Alternatively, in higher dimensions it may be more appropriate to use non-visual techniques, such as hypothesis testing. 

While we have considered ODE models, there is appetite in the literature for parameter estimation, uncertainty quantification and identifiability analysis for more complicated models; including partial differential equations, stochastic differential equations (SDEs), delay differential equations \cite{Bishwal2008,Mehrkanoon2014,Simpson2020}. This appetite extends to non-differential-equation-based models, including agent-based models \cite{Lambert2018} and network models \cite{Hazelton2010}. A natural extension of this work is to present examples demonstrating how the information geometry techniques can be applied to these more complicated models. This will introduce new challenges, though it may be possible to leverage existing techniques; for example, linear noise approximation may be used to obtain a representation of the Fisher information matrix for SDEs \cite{Komorowski2011}. Further, we fix $\sigma$ across observation times, model parameters and populations. However, the techniques presented in this work can be generalised to handle data with non-constant variance \cite{Browning2021b}; the expression for the Fisher information matrix given in Equation (\ref{eq:FIM}) can be extended to account for a parameter-dependent covariance matrix \cite{Malago2015}. Investigation of examples paralleling those in Section \ref{sec:Results}, but with non-constant standard deviation, may prove insightful.        

Here, the Fisher information defines a Riemann metric on the statistical manifold. For some inference problems it is not practical to obtain the Fisher information. Where the Fisher information is not available, the sample-based \textit{observed information}---computed as negative the Hessian of the log-likelihood function, or via Monte Carlo methods---may be available \cite{Efron1978,Poyiadjis2011}. The observed information has been demonstrated to equip a manifold with an \textit{observed geometric structure} akin to the \textit{expected geometric structure} associated with the Fisher information \cite{BarndorffNielsen1986}. Further work could identify the viability of the techniques presented here in situations where only the observed information is available, particularly for local approximation about the MLE.     


\section*{Data accessibility}
Data and code is made available on \href{https://github.com/Jesse-Sharp/Sharp2021b}{GitHub}.

\section*{Competing interests}
We declare we have no competing interests.

\section*{Funding}
J.A.S. and A.P.B. acknowledge support from the Australian Government Research Training Program. J.A.S  acknowledges support from the AF Pillow Applied Mathematics Trust. J.A.S., A.P.B. and K.B. acknowledge support from the Australian Centre of Excellence for Mathematical and Statistical Frontiers (CE140100049). M.J.S. is supported by the Australian Research Council (DP200100177). 

\section*{Acknowledgements}
We thank Professor Alan Garfinkel (UCLA) for introducing K.B. to information geometry, and for numerous discussions while both were at the University of Oxford. We also thank Dr. Brodie Lawson (QUT) for some helpful discussions. \cbl Finally, we thank two anonymous referees for their helpful comments. \cb


\clearpage
\begin{thebibliography}{}
	\bibitem{Amari1985} Amari S. 1985 \textit{Differential-Geometrical Methods in Statistics}. New York: Springer-Verlag.
	
	
	\bibitem{Amari1998} Amari S. 1998 Natural gradient works efficiently in learning. \textit{Neural Computation} \textbf{10}, 251--276. (doi.org/10.1162/089976698300017746).
	
		
	\bibitem{Amari2000} Amari S, Nagaoka H. 2000 \textit{Methods of Information Geometry}. Providence: American Mathematical Society. 
		
	\bibitem{Amari2010} Amari S, Andrzej C. 2010 Information geometry of divergence functions. \textit{Bulletin of the Polish Academy of Sciences. Technical Sciences} \textbf{58}, 183--195. (doi.org/10.2478/v10175-010-0019-1).  
	
	
	\bibitem{Amari2016} Amari S. 2016 \textit{Information Geometry and its Applications}. Tokyo: Springer Japan. 
	
	\bibitem{Andrieu2003} Andrieu C, De Freitas N, Doucet A, Jordan MI. 2003 An introduction to MCMC for machine learning. \textit{Machine Learning} \textbf{50}, 5--43. (doi.org/10.1023/A:1020281327116).
	
	\bibitem{Arutjunjan2020} Arutjunjan R. 2020 On the Geometric Foundation of Parameter Inference. Masters Thesis. Erlangen-N\"{u}rnberg: Friedrich-Alexander University. 
	
	\bibitem{Arwini2008} Arwini KA, Dodson, CTJ. 2008 \textit{Information Geometry -- Near Randomness and Near Independence}. Berlin: Springer-Verlag. 
		

	\bibitem{Audoly2001} Audoly S, Bellu G, D'Angio L, Saccomani MP, Cobelli C. 2001 Global identifiability of nonlinear models of biological systems. \textit{IEEE Transactions on Biomedical Engineering} \textbf{48}, 55--65. (doi.org/10.1109/10.900248).

	\bibitem{BarndorffNielsen1986} Barndorff-Nielsen OE 1986. Likelihood and observed geometries. \textit{The Annals of Statistics} \textbf{14}, 856--873. (doi.org/10.1214/aos/1176350038).
	
	\bibitem{Bezanson2012} Bezanson J, Karpinski S, Shah VB, Edelman A. 2012 Julia: a fast dynamic language for technical computing. \textit{arXiv preprint}. (arxiv.org/abs/1209.5145) 
	
	\bibitem{Bishwal2008} Bishwal JPN. 2008 \textit{Parameter Estimation in Stochastic Differential Equations}. Berlin: Springer-Verlag.
	
	\bibitem{Brouwer2016} Brouwer AF, Meza R, Eisenberg MC. 2016 Parameter estimation for multistage clonal expression models from cancer incidence data: a practical identifiability analysis. \textit{PLoS Computational Biology} \textbf{13}, e1005431. (doi.org/10.1371/journal.pcbi.1005431).
	
	\bibitem{Browning2020} Browning AP, Warne DJ, Burrage K, Baker RE, Simpson MJ. 2020 Identifiability analysis for stochastic differential equation models in systems biology. \textit{Journal of the Royal Society Interface} \textbf{17}, 20200652. (doi.org/10.1098/rsif.2020.0652). 
	
	\bibitem{Browning2021a} Browning AP, Sharp JA, Murphy RJ, Gunasingh G, Lawson B, Burrage K, Haass NK, Simpson MJ. 2021 Quantitative analysis of tumour spheroid structure. \textit{eLife} \textbf{10}, e73020. (doi.org/10.7554/eLife.73020).
	
	\bibitem{Browning2021b} Browning AP, Maclaren OJ, Buenzli PR, Lanaro M, Allenby MC, Woodruff MA, Simpson MJ. Model-based data analysis of tissue growth in thin 3D printed scaffolds. \textit{Journal of Theoretical Biology} \textbf{528}, 110852. (doi.org/10.1016/j.jtbi.2021.110852).
	
	\bibitem{Calin2014} Calin O, Udriste C. 2014 \textit{Geometric Modeling in Probability and Statistics}. Cham: Springer International Publishing. 
	
	
	\bibitem{Calvo1991} Calvo M, Oller JM. 1991 An explicit solution of information geodesic equations for the multivariate normal model. \textit{Statistics \& Risk Modeling} \textbf{9}, 119--138. (doi.org/10.1524/strm.1991.9.12.119).
	
	\bibitem{Carcione2020} Carcione JM, Santos JE, Bagaini C, Ba J. 2020 A simulation of a covid-19 epidemic based on a deterministic SIR model. \textit{Frontiers in Public Health} \textbf{8}, 1--13. (doi.org/10.3389/fpubh.2020.00230). 
	
	\bibitem{Costa2015} Costa SIR, Santos SA, Strapasson JE. 2015 Fisher information distance: a geometrical reading. \textit{Discrete Applied Mathematics} \textbf{197}, 59--69. (doi.org/10.1016/j.dam.2014.10.004). 
	
	\bibitem{Daly2018} Daly AC, Gavaghan D, Cooper J, Tavener S. 2018 Inference-based assessment of parameter identifiability in nonlinear biological models. \textit{Journal of the Royal Society Interface} \textbf{15}, 20180318. (doi.org/10.1098/rsif.2018.0318). 
	
	\bibitem{DevenishNelson2010} Devenish--Nelson ES, Harris S, Soulsbury CD, Richards SA, Stephens PA. Uncertainty in population growth rates: determining confidence intervals from point estimates of parameters. \textit{PLOS ONE} \textbf{5}, e13628. (doi.org/10.1371/journal.pone.0013628). 
	
	\bibitem{Dobrushkin2017} Dobrushkin VA. 2017 \textit{Applied Differential Equations with Boundary Value Problems}. Boca Raton: CRC Press.
	
	
	\bibitem{EdelsteinKeshet2005} Edelstein-Keshet L. 2005 \textit{Mathematical Models in Biology}. Philadelphia: Society for Industrial and Applied Mathematics. 
	
	\bibitem{Efron1978} Efron B, Hinkley DV. 1978 Assessing the accuracy of the maximum likelihood estimator: Observed versus expected Fisher information. \textit{Biometrika} \textbf{65}, 457--487. (doi.org/10.1093/biomet/65.3.457). 
	
	\bibitem{Eguchi2006} Eguchi S, Copas J. 2006 Interpreting Kullback-Leibler divergence with the Neyman-Pearson lemma. \textit{Journal of Multivariate Analysis} \textbf{97}, 2034--2040. (doi.org/10.1016/j.jmva.2006.03.007).
	
	\bibitem{Ferson1996} Ferson S, Ginzburg LR. 1996 Different methods are needed to propagate ignorance and variability. \textit{Reliability Engineering and System Safety} \textbf{54}, 133--144. (doi.org/10.1016/S0951-8320(96)00071-3).
	
	\bibitem{Fisher1992} Fisher RA. 1992 \textit{Statistical methods for research workers} In: Kotz S, Johnson NL (eds) Breakthroughs in Statistics, 66--70. New York: Springer.
	
	\bibitem{Frieden1990} Frieden BR. 1990 Fisher information, disorder, and the equilibrium distribution of physics. \textit{Physical Review A} \textbf{41}, 4265--4276. (doi.org/10.1103/PhysRevA.41.4265).
	
	\bibitem{Frieden2007} Frieden BR, Gatenby RA. 2007 \textit{Exploratory Data Analysis Using Fisher Information}. London: Springer London. 
	
	\bibitem{Frolich2014} Fr\"{o}hlich F, Theis FJ, Hasenauer J. 2014 Uncertainty analysis for non-identifiable dynamical systems: profile likelihoods, bootstrapping and more. In \textit{Computational Methods in Systems Biology: Proceedings of the 12th International Conference, 2014}. Cham: Springer.
	
	\bibitem{Gabor2015} G\'{a}bor A, Banga JR. 2015 Robust and efficient parameter estimation in dynamic models of biological systems. \textit{BMC Systems Biology} \textbf{9}, 1--25. (doi.org/10.1186/s12918-015-0219-2). 
	
	\bibitem{Gelman1995} Gelman A, Carlin JB, Stern HS, Dunson SB, Vehtari A, Rubin DB. 1995 \textit{Bayesian data analysis}. New York: Chapman \& Hall/CRC. 
	
	\bibitem{Gelman1996} Gelman A, Bois F, Jiang J. 1996 Physiological pharmacokinetic analysis using population modeling and informative prior distributions. \textit{Journal of the American Statistical Association} \textbf{91}, 1400--1412. (doi:10.2307/2291566). 
	
	\bibitem{Giesel2021} Giesel E, Reischke R, Sch{\"a}fer BM, Chia D. 2021 Information geometry in cosmological inference problems. \textit{Journal of Cosmology and Astroparticle Physics} \textbf{2021}, 005. (doi/org/10.1088/1475-7516/2021/01/005). 
	
	\bibitem{Girolami2011} Girolami M, Calderhead B. 2011 Riemann manifold Langevin and Hamiltonian Monte Carlo methods. \textit{Journal of the Royal Statistical Society B} \textbf{73}, 123–-214. (doi.org/10.1111/j.1467-9868.2010.00765.x).
	
	\bibitem{Godfrey1980} Godfrey KR, Jones RP, Brown RF. 1980 Identifiable pharmacokinetic models: the role of extra inputs and measurements. \textit{Journal of pharmacokinetics and biopharmaceutics} \textbf{8}, 633--648. (doi.org/10.1007/bf01060058).
	
	\bibitem{Hagen2013} Hagen DR, White JK, Tidor B. 2013 Convergence in parameters and predictions using computational experimental design. \textit{Interface Focus} \textbf{3}, 20130008. (doi.org/10.1098/rsfs.2013.0008).
	
	\bibitem{Hay2021} Hay J, Hellewell J, Qiu X. 2021 When intuition falters: repeated testing accuracy during an epidemic. \textit{European Journal of Epidemiology} \textbf{36}, 749--752. (doi.org/10.1007/s10654-021-00786-w).
	
	\bibitem{Hazelton2010} Hazelton ML. 2010 Bayesian inference for network-based models with a linear inverse structure. \textit{Transportation Research Part B: Methodological} \textbf{44}, 674--685. (doi.org/10.1016/j.trb.2010.01.006). 
	
	\bibitem{Heinbockel2001} Heinbockel JH. 2001 \textit{Introduction to Tensor Calculus and Continuum Mechanics}. Victoria, BC: Trafford. 
	
	\bibitem{Hines2014} Hines KE, Middendorf TR, Aldrich RW. 2014 Determination of parameter identifiability in nonlinear biophysical models: a Bayesian approach. \textit{Journal of General Physiology} \textbf{143}, 401--416. (doi.org/10.1085/jgp.201311116). 
	
	\bibitem{Hlavacek2011} Hlavacek WS. 2011 \textit{Two Challenges of Systems Biology. In Handbook of Statistical Systems Biology}. Chichester: John Wiley \& Sons, Ltd.
	
	\bibitem{House2016} House T, Ford A, Lan S, Bilson S, Buckingham-Jeffery E, Girolami M. 2016 Bayesian uncertainty quantification for transmissibility of influenza, norovirus and Ebola using information geometry. \textit{Journal of the Royal Society Interface} \textbf{13}, 20160279. (doi.org/10.1098/rsif.2016.0279).  
	
	\bibitem{Jabot2013} Jabot F, Faure T, Dumoulin N. 2013 EasyABC: performing efficient approximate Bayesian computation sampling schemes in R. \textit{Methods in Ecology and Evolution} \textbf{4}, 684--687. (doi.org/ 10.1111/2041-210X.12050).
	
	\bibitem{Jost2017} Jost J. 2017 \textit{Riemannian Geometry and Geometric Analysis}. Cham: Springer International Publishing. 
	
	
	\bibitem{Kass1997} Kass RE, Vos PW. 1997 \textit{Geometrical Foundations of Asymptotic Inference}. New Jersey: Wiley-Interscience.
	
	\bibitem{Kay1993} Kay SM. 1993 \textit{Fundamentals of Statistical Signal Processing: Estimation Theory}. New Jersey: Prentice-Hall. 
	
	\bibitem{Kermack1927} Kermack WO, McKendrick AG. 1927 A contribution to the mathematical theory of epidemics. \textit{Proceedings of the Royal Society A} \textbf{115}, 700--721. (doi.org/10.1098/rspa.1927.0118). 
	
	\bibitem{Komorowski2011} Komorowski M, Costa MJ, Rand DA, Stumpf MPH. 2011 Sensitivity, robustness and identifiability in stochastic chemical kinetics models. \textit{Proceedings of the National Academy of Sciences of the United States of America} \textbf{108}, 8645--8650. (doi.org/10.1073/pnas.1015814108).
	
	\bibitem{Kursawe2018} Kursawe J, Baker RE, Fletcher AG. Approximate Bayesian computation reveals the importance of repeated measurements for parameterising cell-based models of growing tissues. \textit{Journal of Theoretical Biology} \textbf{443}, 66--81. (doi.org/10.1016/j.jtbi.2018.01.020).   
	
	\bibitem{Lambert2018} Lambert B, MacLean AL, Fletcher AG, Combes AN, Little MH, Byrne HM. 2018 Bayesian inference of agent-based models: a tool for studying kidney branching morphogenesis. \textit{Journal of Mathematical Biology} \textbf{76}, 1673--1697). (doi.org/10.1007/s00285-018-1208-z).
	
	\bibitem{Lambert2018b} Lambert B. 2018\textit{ A Student's Guide to Bayesian Statistics}. London: Sage Publications Ltd.
	
	\bibitem{Lee2018} Lee JM. 2018 \textit{Introduction to Riemannian Manifolds}. Cham: Springer International Publishing. 
	
	
	\bibitem{Lehmann1998} Lehmann EL, Fienberg S, Casella G. 1998 \textit{Theory of Point Estimation}. Secaucus: Springer. 
	
	\bibitem{Liepe2010} Liepe J, Barnes C, Cule E, Erguler K, Kirk P, Toni T, Stumpf MPH. 2010 ABC-SysBio--approximate Bayesian computation in Python with GPU support. \textit{Bioinformatics} \textbf{26}, 1797--1799. (doi.org/10.1093/bioinformatics/btq278).
	
	\bibitem{Liepe2013} Liepe J, Filippi S, Komorowski M, Stumpf MPH. 2013 Maximising the information content of experiments in systems biology. \textit{PLOS Computational Biology} \textbf{9}, e1002888. (doi.org/10.1371/journal.pcbi.1002888.g002).   
	
	\bibitem{Liepe2014} Liepe J, Kirk P, Filippi S, Toni T, Barnes CP, Stumpf MPH. 2014 A framework for parameter estimation and model selection from experimental data in systems biology using approximate Bayesian computation. \textit{Nature Protocols} \textbf{9}, 439--456. (doi.org/10.1038/nprot.2014.025).
	
	
	\bibitem{Lill2019} Lill D, Timmer J, Kaschek D. 2019 Local Riemannian geometry of model manifolds and its implications for practical parameter identifiability. \textit{PLOS ONE}, \textbf{14}: e0217837. (doi.org/10.1371/journal.pone.0217837). 
	
	\bibitem{Liu1998} Liu JS, Chen R. 1998 Sequential Monte Carlo methods for
	dynamic systems. \textit{Journal of the American Statistical Association}, \textbf{93}, 1032--1044. (10.1080/01621459.1998.10473765).
	
	\bibitem{Loveridge2016} Loveridge LC. 2016 Physical and geometric interpretations of the Riemann Tensor, Ricci Tensor and Scalar Curvature. \textit{arXiv preprint}. (arxiv.org/abs/gr-qc/0401099).
	
	\bibitem{Luscombe2018} Luscombe JH. 2018 \textit{Core Principles of Special and General Relativity}. Boca Raton: CRC Press, Taylor \& Francis.
	
	\bibitem{Maclaren2020} Maclaren OJ, Nicholson R. 2020 What can be estimated? Identifiability, estimability, causal
	inference and ill-posed inverse problems. \textit{arXiv preprint}. (arxiv.org/abs/1904.02826).
	
	\bibitem{Malago2015} Malag{\`o} L, Pistone G. 2015 Information geometry on the Gaussian distribution in view of stochastic optimization. \textit{ Proceedings of the 2015 ACM Conference on Foundations of Genetic Algorithms XIII} \textbf{13}, 150--162. (doi.org/10.1145/2725494.2725510).
	
	\bibitem{Marino2008} Marino S, Hogue IB, Ray CJ, Kirschner DE. 2008 A methodology for performing global uncertainty analysis in systems biology. \textit{Journal of Theoretical Biology} \textbf{254}, 178--196. (doi.org/10.1016/j.jtbi.2008.04.011).
	
	\bibitem{Marjoram2003} Marjoram P, Molitor J, Plagnol V, Tavar{\'e} S. 2013 Markov chain Monte Carlo without likelihoods. \textit{Proceedings of the National Academy of Sciences of the United States of America} \textbf{100}, 15324--15328. (doi.org/10.1073pnas.0306899100). 
	
	\bibitem{Mehrkanoon2014} Mehrkanoon S, Mehrkanoon S, Suykens JAK. 2014 Parameter estimation of delay differential equations: an integration-free LS-SVM approach. \textit{Communications in Nonlinear Science and Numerical Simulation} \textbf{19}, 830--841. (doi.org/10.1016/j.cnsns.2013.07.024).
	
	\bibitem{Menendez1995} Men{\'e}ndez ML, Morales D, Pardo L, Salicr\'{u} M. 1995 Statistical tests based on geodesic distances. \textit{Applied Mathematics Letters} \textbf{8}, 65--69. (doi.org/10.1016/0893-9659(94)00112-P).
	
	\bibitem{Miao2011} Miao H, Xia X, Perelson AS, Wu H. 2011 On identifiability of nonlinear models and applications in viral dynamics. \textit{SIAM Review} \textbf{53}, 3--39. (doi.org/10.1137/090757009).
	
	\bibitem{Mitra2019} Mitra ED, Hlavacek WS. 2019 Parameter estimation and uncertainty quantification for systems biology models. \textit{Current Opinion in Systems Biology} \textbf{18}, 9--18. (doi.org/10.1016/j.coisb.2019.10.006).
	
	
	\bibitem{Murphy2012} Murphy KP. 2012 \textit{Machine Learning: A Probabilistic Perspective}. Massachusetts: MIT Press.
	
	
	\bibitem{Murray2002} Murray JD. 2002\textit{ Mathematical Biology I: An Introduction, 3rd ed}. Heidelberg: Springer.
	
	\bibitem{Nielsen2020} Nielsen F. 2020 An elementary introduction to information geometry. \textit{Entropy} \textbf{22}, 1--61. (doi.org/10.3390/e22101100). 
	
	\bibitem{Panik2014} Panik MJ. 2014 \textit{Growth Curve Modeling: Theory and Applications}. New Jersey: John Wiley \& Sons, Ltd.
	
	\bibitem{Papst2019} Papst I, Earn DJD. 2019 Invariant predictions of epidemic patterns from radically different forms of seasonal forcing. \textit{Journal of the Royal Society Interface} \textbf{16}, 20190202. (doi.org/10.1098/rsif.2019.0202).
	
	\bibitem{Pawitan2001} Pawitan Y. 2001 \textit{In All Likelihood: Statistical Modelling and Inference Using Likelihood}. Oxford: Oxford University Press, Incorporated.  
	
	
	\bibitem{Pinele2019} Pinele J, Costa SIR, Strapasson JE. 2019 \textit{On the Fisher-Rao Information Metric in the Space of Normal Distributions}. Cham: Springer International Publishing AG.    
	
	
	\bibitem{Pinele2020} Pinele J, Strapasson JE, Costa SIR. 2020 The Fisher-Rao distance between multivariate normal distributions: special cases, bounds and applications. \textit{Entropy} \textbf{22}, 1--24. (doi.org/10.3390/e22040404).
	
	\bibitem{Poyiadjis2011} Poyiadjis G. 2011 Particle approximations of the score and observed information matrix in state space models with application to parameter estimation. \textit{Biometrika} \textbf{98}, 65--80. (doi.org/10.1093/biomet/asq062).
	
	\bibitem{Powell2009} Powell MJD. 2009 The BOBYQA algorithm for bound constrained optimization without derivatives. \textit{Technical Report DAMTP 2009/NA06, Department of Applied Mathematics and Theoretical Physics, University of Cambridge}. (https://www.damtp.cam.ac.uk/user/na/NA\_papers/NA2009\_06.pdf).  
		
	\bibitem{Prangle2017} Prangle D. 2017 Adapting the ABC distance function. \textit{Bayesian Analysis} \textbf{12}, 289--309. (doi.org/10.1214/16-BA1002).

	
	 \bibitem{Rackauckas2017} Rackauckas C, Nie Q. 2017 DifferentialEquations.jl – A Performant and Feature-Rich Ecosystem for Solving Differential Equations in Julia. \textit{The Journal of Open Research Software} \textbf{5}, 1--10. (doi.org/10.5334/jors.151). 
		
	\bibitem{Rao1992} Rao CR. 1992 \textit{Information and the Accuracy Attainable in the Estimation of Statistical Parameters.} In: Kotz S, Johnson NL (eds) Breakthroughs in Statistics, 235--247. New York: Springer.
	
	\bibitem{Raue2009} Raue A, Kreutz C, Maiwald T, Bachmann J, Schilling M, Klingm{\"u}ller U, Timmer J. 2009 Structural and practical identifiability analysis of partially observed dynamical models by exploiting the profile likelihood. \textit{Bioinformatics} \textbf{25}, 1923--1929. (doi.org/10.1093/bioinformatics/btp358).
	
	\bibitem{Ruess2015} Ruess J, Parise F, Milias-Argeitis A, Khammash M, Lygeros J. 2015 Iterative experiment design guides the characterization of a light-inducible gene expression circuit. \textit{Proceedings of the National Academy of Sciences of the United States of America} \textbf{112}, 8148--8153. (doi.org/10.1073/pnas.1423947112).  
	
	\bibitem{Roberts2007} Roberts MG. 2007 The pluses and minuses of $\mathcal{R}_0$. \textit{Journal of the Royal Society Interface} \textbf{4}, 949--961. (doi.org/10.1098/rsif.2007.1031).  
	
	\bibitem{Schnoerr2017} Schnoerr D, Sanguinetti G, Grima R. 2017 Approximation and inference methods for stochastic biochemical kinetics -- a tutorial review. \textit{Journal of Physics A: Mathematical and Theoretical} \textbf{50}, 093001. (doi.org/10.1088/1751-8121/aa54d9).
	
	\bibitem{Seghouane2007} Seghouane A, Amari S. 2007 The AIC criterion and symmetrizing the Kullback–Leibler divergence. \textit{IEEE Transactions on Neural Networks} \textbf{18}, 97--106. (doi.org/10.1109/TNN.2006.882813).   
	
	\bibitem{Sharp2020} Sharp JA, Browning AP, Mapder T, Baker CM, Burrage K, Simpson MJ. 2020 Designing combination therapies using multiple optimal controls. \textit{Journal of Theoretical Biology} \textbf{497}, 110277. (doi.org/10.1016/j.jtbi.2020.110277).
	
	\bibitem{Siekmann2012} Siekmann I, Sneyd J, Crampin EJ. 2012 MCMC can detect nonidentifiable models. \textit{Biophysical Journal} \textbf{103}, 2275--2286. (doi.org/10.1016/j.bpj.2012.10.024).  
	
	\bibitem{Simpson2020} Simpson MJ, Baker RE, Vittadello ST, Maclaren OJ. 2020 Practical parameter identifiability for	spatio-temporal models of cell invasion. \textit{Journal of the Royal Society Interface} \textbf{17}, 20200055. (doi.org/10.1098/rsif.2020.0055).
	
	\bibitem{Simpson2021} Simpson MJ, Browning AP, Warne DJ, Maclaren OJ, Baker RE. 2021 Parameter identifiability and model selection for sigmoid population growth models. \textit{Journal of Theoretical Biology} \textbf{535}, 110998. (doi.org/10.1016/j.jtbi.2021.110998). 
	
	\bibitem{Sisson2018} Sisson SA, Fan Y, Beaumont M. 2018 \textit{Handbook of Approximate Bayesian Computation}. Boca Raton: Chapman \& Hall-CRC, Taylor and Francis Group. 
	
	\bibitem{Sunnaker2013} Sunn$\mathring{\textrm{a}}$ker M, Busetto AG, Numminen E, Corander J, Foll M, Dessimoz C. 2013 Approximate Bayesian computation. \textit{PLOS Computational Biology} \textbf{9}, e1002803. (doi.org/10.1371/journal.pcbi.1002803).
	
	\bibitem{Toni2009} Toni T, Welch D, Strelkowa N, Ipsen A, Stumpf MPH. 2009 Approximate Bayesian computation scheme for parameter inference and model selection in dynamical systems. \textit{Journal of the Royal Society Interface} \textbf{6}, 187--202. (doi.org/10.1098/rsif.2008.0172).
	
	\bibitem{Transtrum2014} Transtrum MK, Qiu P. 2014 Model reduction by manifold boundaries. \textit{Physical Review Letters} \textbf{113}, 098701. (doi.org/10.1103/PhysRevLett.113.098701).
	
	\bibitem{Tsitouras2011} Tsitouras C. 2011 Runge–Kutta pairs of order 5 (4) satisfying only the first column simplifying assumption. \textit{Computers \& Mathematics with Applications} \textbf{62}, 770--775. (doi.org/10.1016/j.camwa.2011.06.002).
	
	\bibitem{Tsoularis2002} Tsoularis A, Wallace J. 2002 Analysis of logistic growth models. \textit{Mathematical Biosciences} \textbf{179}, 21--55. (doi.org/10.1016/S0025-5564(02)00096-2).

	\bibitem{Villaverde2016} Villaverde AF, Barreiro A, Papachristodoulou A. 2016 Structural identifiability of dynamic systems biology models. \textit{PLOS Computational Biology} \textbf{12}, e1005153. (doi.org/10.1371/journal.pcbi.1005153).

	\bibitem{Villaverde2019} Villaverde AF. 2019 Observability and structural identifiability of nonlinear biological systems. \textit{Complexity} \textbf{2019}, 1--12. (doi.org/10.1155/2019/8497093).
	
	\bibitem{Vo2015} Vo BN, Drovandi CC, Pettitt AN, Simpson MJ. 2015 Quantifying uncertainty in parameter estimates for stochastic models of collective cell spreading using approximate Bayesian computation. \textit{Mathematical Biosciences} \textbf{263}, 133--142. (doi.org/10.1016/j.mbs.2015.02.010).
	
	\bibitem{Walpole2013} Walpole J, Papin JA, Peirce SM. 2013 Multiscale computational models of complex biological systems. \textit{Annual Review of Biomedical Engineering} \textbf{15}, 137--154. (doi.org/10.1146/annurev-bioeng-071811-150104). 
	
	\bibitem{Warne2017} Warne DJ, Baker RE, Simpson MJ. 2017 Optimal quantification of contact inhibition in cell populations. \textit{Biophysical Journal} \textbf{113}, 1920--1924. (doi.org/10.1016/j.bpj.2017.09.016).
	
	\bibitem{Warne2019} Warne DJ, Baker RE, Simpson MJ. 2019 Simulation and inference algorithms for stochastic biochemical reaction networks: from basic concepts to state-of-the-art. \textit{Journal of the Royal Society Interface} \textbf{16}, 20180943. (doi.org/10.1098/rsif.2018.0943). 
	
	\bibitem{Warne2019a} Warne DJ, Baker RE, Simpson MJ. 2019 Using experimental data and information criteria to guide model selection for reaction--diffusion problems in mathematical biology. \textit{Bulletin of Mathematical Biology} \textbf{81}, 1760--1804. (doi.org/10.1007/s11538-019-00589-x).
	
	\bibitem{Warne2020} Warne DJ, Baker RE, Simpson MJ. 2020 A practical guide to pseudo-marginal methods for computational inference in systems biology. \textit{Journal of Theoretical Biology} \textbf{496}, 110255. (doi.org/10.1016/j.jtbi.2020.110255). 
	
	\bibitem{Watanabe2009} Watanabe S. 2009 \textit{Algebraic Geometry and Statistical Learning Theory}. New York: Cambridge University Press.
	
	\bibitem{Wilkinson2013} Wilkinson RD. 2013 Approximate Bayesian computation (ABC) gives exact results under the assumption of model error. \textit{Statistical Applications in Genetics and Molecular Biology} \textbf{12}, 129--141. (doi.org/10.1515/sagmb-2013-0010).
	
	\bibitem{Yeung2020} Yeung E, McFann S, Marsh L, Dufresne E, Filippi S, Harrington HA, Shvartsman SY, W{\"u}hr M. 2020 Inference of multisite phosphorylation rate constants and their modulation by pathogenic mutations. \textit{Current Biology} \textbf{30}, 877--882. (doi.org/10.1016/j.cub.2019.12.052).
	
	\bibitem{Yue2006} Yue H, Brown M, Knowles J, Wang H, Broomhead DS, Kell DB. Insights into the behaviour of systems biology models from dynamic sensitivity analysis: a case study of an NF-$\kappa$B signalling pathway. \textit{Molecular Biosystems} \textbf{2}, 640--649. (doi.org/10.1039/B609442B).
	
	\bibitem{Zimmer2018} Zimmer C, Leuba SI, Cohen T, Yaesoubi R. 2018 Accurate quantification of uncertainty in epidemic parameter estimates and predictions using stochastic compartmental models. \textit{Statistical Methods in Medical Research} \textbf{28}, 3591--3608. (doi.org/10.1177/0962280218805780).  
\end{thebibliography}
\end{document}